\documentclass[runningheads]{llncs}
\usepackage[T1]{fontenc} 
\usepackage{graphicx}
\usepackage{xcolor}
\usepackage{algorithm}
\usepackage{algpseudocode}
\usepackage{color,soul}
\usepackage{adjustbox, soul, color}
\begin{document}
\title{When the Guard failed the Droid: A case study of Android malware}
%
%
\author{Harel Berger\inst{1} \and
Chen Hajaj\inst{2} \and
Amit Dvir\inst{3}
}
\authorrunning{H. Berger et al.}
%
\institute{Ariel Cyber Innovation Center, CS Department, Ariel University, Kiryat Hamada Ariel 40700,  Israel\\
\email{harel.berger@msmail.ariel.ac.il}\\
\and
Ariel Cyber Innovation Center, Data Science and Artificial Intelligence Research Center, IEM Department, Ariel University,  Israel\\
\email{chenha@ariel.ac.il}\\
\url{https://www.ariel.ac.il/wp/chen-hajaj/}
\and
Ariel Cyber Innovation Center, CS Department, Ariel University,  Israel\\
\email{amitdv@ariel.ac.il}\\
\url{https://www.ariel.ac.il/wp/amitd/} 
}
\maketitle              
\begin{abstract}
Android malware is a persistent threat to billions of users around the world. As a countermeasure, Android malware detection systems are occasionally implemented. However, these systems are often vulnerable to \emph{evasion attacks}, in which an adversary manipulates malicious instances so that they are misidentified as benign. In this paper, we launch various innovative evasion attacks against several Android malware detection systems. The vulnerability inherent to all of these systems is that they are part of Androguard~\cite{desnos2011androguard}, a popular open source library used in Android malware detection systems. Some of the detection systems decrease to a 0\% detection rate after the attack. Therefore, the use of open source libraries in malware detection systems calls for caution. 

In addition, we present a novel evaluation scheme for evasion attack generation that exploits the weak spots of known Android malware detection systems. In so doing, we evaluate the functionality and maliciousness of the manipulated instances created by our evasion attacks. We found variations in both the maliciousness and functionality tests of our manipulated apps. We show that non-functional apps, while considered malicious, do not threaten users and are thus useless from an attacker's point of view. We conclude that evasion attacks must be assessed for both functionality and maliciousness to evaluate their impact, a step which is far from commonplace today. 

\end{abstract}
\pagebreak
\section{Introduction}
\label{Intro}
Malware, or malicious software, is defined as any program or file that is harmful to a computer user. Malware can appear in many forms, such as viruses, Trojans, and worms. Among others, the Android OS is susceptible to this threat as well.
Android malware evolves over time and take many new forms, such that malicious versions of the popular Android application PacKages (APKs) can propagate to various Android markets. For example, Xhelper~\cite{Xhelper} was recently shown to be sophisticated Android malware. It is able reinstall itself after the uninstall process. In addition, it is designed to stay hidden as it does not create an icon at the launcher. This app has infected over 45k devices in six months. It does not require dangerous permissions from the user but instead operates based on device events such as completion of the boot process. It incorporates persistent connection to a C \& C server to send actions to the device. In advance, it was discovered by MalwareByte Labs~\cite{Xhelper_adv} that some variations of Xhelper may like the host so much that it may reinstall itself even after a factory reset of the phone. The source of installation for the malware stated it was coming from Google play. It is believed that something inside Google play triggers the quick installation of the app. It is still considered a mystery.

Not all Android malware are as complex as Xhelper. But based on this advanced example and many others, a long list of researchers have reached the inevitable conclusion that Android malware detection is a major issue that should be tackled by the security community~\cite{arp2014drebin,cai2018towards,chen2016stormdroid,dini2012madam,huynh2017new,onwuzurike2019mamadroid,shabtai2012andromaly,shabtai2014mobile,wu2012droidmat}. As Android malware detection is a subdomain of malware detection, general malware detection methods apply to Android malware as well. One of the most popular techniques in the malware detection domain is Machine Learning (ML) based detection of malicious entities~\cite{bergeron2001static,laskov2011static,nath2014static,shabtai2009detection,vsrndic2016hidost}, where some techniques' detection rates exceed 99\% at times~\cite{laskov2013detection,laskov2014practical}. 

Machine Learning (ML) refers to a set of algorithms and statistical models that systems use to perform tasks by relying on patterns and inference instead of explicit instructions~\cite{mcdonald1989machine}. ML algorithms build mathematical models based on sample data, also known as training data, to make predictions or decisions on other sample data, that was not previously observed before. The typical approach in ML is to extract a set of features, or attributes, from each training sample and learn a model that labels the samples, which are presented in terms of the former extracted features. In the case of malware detection, these algorithms need to distinguish between malicious and benign instances, hence the focus on classification tasks with a binary target variable. 

However, Szegedy et al. \cite{goodfellow2014explaining} showed that some ML methods (including malware detection systems) are vulnerable to~\textit{adversarial examples}. Adversarial examples (AE) are samples an attacker manipulates, such that the sample will be wrongly classified (e.g., malicious instances as benign). These examples are manipulated such that they hide some properties, or adopt properties of a different class, such that the system misclassifies them. In the context of malware detection, misclassification involves classifying benign samples as malicious and vice-versa \cite{grosse2017adversarial,kuppa2019black,yuan2019adversarial}. AE attacks are widely used in the many fields: spam filtering~\cite{nelson2008exploiting}, network intrusion detection systems (IDS)~\cite{fogla2006polymorphic}, Anti-Virus signature tests~\cite{newsome2006paragraph} and  biometric recognition~\cite{rodrigues2009robustness}. 

There are three basic types of attackers generating AE: the white-box attacker~\cite{madry2017towards,raghunathan2018certified,song2017pixeldefend}  which has full knowledge of the classifier and the train/test data and the black-box attacker~\cite{madry2017towards,papernot2017practical,shahpasand2019adversarial}, whose number of queries to an oracle, access to the predicted probabilities or the predicted classes from the model, or access to the training data are limited. Finally, the zero-knowledge attacker~\cite{goldreich1986prove,kurakin2016adversarial}, has no prior data before the attack.
\space A special case of adversarial examples involves using  \textit{evasion attacks}. Evasion attacks take place when an adversary modifies malware code so that the modified malware is categorized as benign by the ML, but still successfully executes the malicious payload~\cite{fogla2006polymorphic,grosse2016adversarial,laskov2014practical,maiorca2013looking,xu2016automatically}. In Android malware detection systems, various studies have explored AE-based attack methods~\cite{chen2018android,grosse2016adversarial,grosse2017statistical,article_liu}. On the other hand, AE techniques have also been used to robustify classifiers against Android malware~\cite{chen2017securedroid,grosse2016adversarial,Hu2017}. 

The main contribution of this work is introducing a vulnerability in one of well known components of Android malware detection systems, Androguard~\cite{desnos2011androguard}, and exploiting it to allow malicious instances to be classified as benign. Androguard implements a  basic feature extraction for Android malware detection systems, and may seem safe. However, since the Android OS continuously updates its APIs, and due to the fact that Androguard has not been updated in this manner, it becomes a weakspot for large numbers of Android malware detection system that uses Androguard~\cite{alazab2020intelligent,gonzalez2020development,ma2020droidetec,raghuraman2020static,sharmeen2020adaptive}. Given this security breach, we design various novel evasion attacks against Android malware detection systems based on Androguard.  

We did not solely focus our analysis on the evasion rate (i.e., the rate of malicious instances bypassing the classifier), as commonly done, but rather conducted a more extensive analysis. Specifically, a closer look at the manipulated instances of the evasion attacks revealed that some instances maintain high evasion and maliciousness rates but do not pass a simplistic functionality test. Based on this analysis, we conclude that further research in Android malware field should consider the impact of evasion attacks in terms of: functionality, maliciousness, and evasion rate.
\section{Background}
The Android OS is an open source OS, mainly based on Linux, and considered one of the most popular mobile phone operating systems. Android was developed by Google~\cite{google_con} and a group of developers known as the Open Handset Alliance~\cite{op_hand}. The distribution of applications for this OS is basically done through application stores such as the Google Play Store~\cite{GooglePlay} or the Samsung Galaxy Store~\cite{galaxy_app}, etc. In contrast to IOS, Android allows independent installation of apps hence leaving itself open to potentially harmful ones. This makes the need for Android malware detection systems even greater. The apps in these markets are stored in an Android PacKage (APK) format. 

\subsection{APK}
\label{apk infra}
\begin{figure}[!]
	\centering
	\includegraphics[width=1
	\columnwidth]{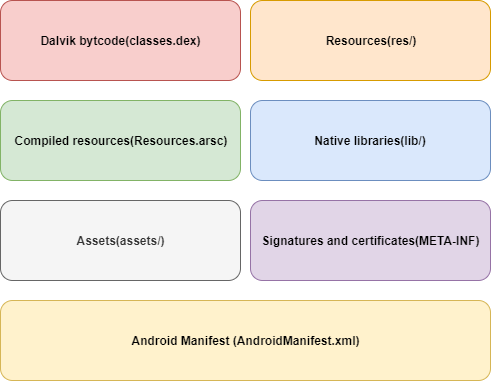}
	\caption{APK structure}
	\label{fig:struct}
\end{figure}

An APK file is a compressed file that
packs the app’s Dalvik bytecode (Classes.dex files), compiled and plain resources, assets, and the XML manifest file as depicted in Fig. \ref{fig:struct}. Dalvik bytecode (Classes.dex) files are the source code that the app runs. The Resources folder contains graphics, string resources, user interface layouts, etc. and complied resources is a binary file that contains the contents of the resources. The native libraries include the outer code libraries that are needed for the app and the Assets folder includes other outer files that are not code files. The signatures and certificates of the apps are also included within the APK. The   certificate and signatures are used to identify the author of the app and establish trust relationships between applications.  
The AndroidManifest.xml (or Manifest file) is designed for the meta-data such as requests for permissions, 
components defined in the app such as Activities, Services, etc.

One of the tags in the Manifest file is \textbf{uses-permission}, which specifies a system permission the user has to grant for the app to operate correctly. Permissions are granted by the user when the app is installed (on Android version 5.1 and lower) or while the app is running (starting from Android version 6.0). An example of this tag is shown in the first box in Fig.~\ref{fig:permission_tag}. A custom tag exists for developer-specific permissions which is called \textbf{permission}. A newer version (on Android version 6.0$+$) of the uses-permission tag is the \textbf{uses-permission-sdk-23} (the second box of Fig. \ref{fig:permission_tag} depicts the use of the new uses-permission tag). Starting from version 1.0, the Android system defined  \textbf{Protection Levels} for each permission: Dangerous, Normal, Signature and Special~\cite{protection_level}. From Android version 6.0$+$, \textbf{Permission Groups} were introduced. These groups put a number of sub-permissions requests together in the same permission tag~\cite{protection_groups} (the third box in Fig. \ref{fig:permission_tag}).

\begin{figure}[!h]
    \centering
    \adjustbox{margin=1em,width=14cm,height=0.8cm,frame,center}{<uses-permission android:name="android.permission.WRITE\_EXTERNAL\_STORAGE"/>}
    \adjustbox{margin=1em,width=14cm,height=0.8cm,frame,center}{<uses-permission-sdk-23 android:name="android.permission.WRITE\_EXTERNAL\_STORAGE"/>}
    \adjustbox{margin=1em,width=14cm,height=0.8cm,frame,center}{<uses-permission-sdk-23 android:name="android.permission.STORAGE"/>}
    \caption{Examples of permission tags. Uppermost: the old uses-permission tag. Middle: the newer version. Bottom: the use of permission groups.
    }
    \label{fig:permission_tag}
\end{figure}
A general tag included in the Manifest file is the \textbf{Meta-data tag}. This is a name-value pair for an item in the Manifest file, which contains additional data that can be supplied to the parent component. A component can contain any number of $<$meta-data$>$ sub-elements, where each meta-data tag includes a name field and a value (which are not constrained), as depicted in Fig. \ref{fig:meta_tag}.
\begin{figure}[!h]
    \centering
    \adjustbox{margin=1em,width=14cm,height=0.8cm,frame,center}{<meta-data android:name="zoo" android:value="@string/kangaroo" />}
    \caption{Example of a meta-data tag}
    \label{fig:meta_tag}
\end{figure}
\begin{figure}[h!]
	\centering
	\includegraphics[width=0.7 \columnwidth]{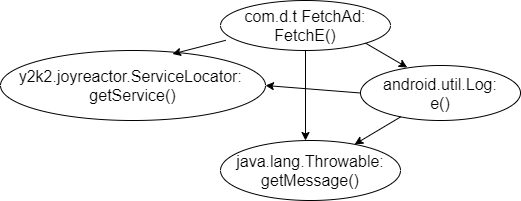}
	\caption{Control Flow Graph. The API call FetchE() calls getService(), getMessage() and the function e from the log class. The e() function calls API calls getService() and getMessage().}
	\label{fig:cfg}
\end{figure}

\subsection{Understanding the APK Code}
\label{under_apk}
As mentioned above, the APK compressed file includes bytecode dex files and a number of other files, as shown in Fig. \ref{fig:struct}. To produce these bytecode files, we use the APKTool~\cite{winsniewski2012android} on the APK compressed file. This tool decompresses an APK file into its subordinate files (and can also be used to compress these files into an APK). These bytecode files are not readable or easily modified by developers. We overcame this by implementing a disassembled code of the app's bytecode code, using the most common disassembled code format for Android,  Smali~\cite{Smali_functions,Smali_github_implementation}. 

The disassembled code is composed of functional code along with various API calls to the device components, such as Camera, GPS, Internet, etc. Each API call is abstracted to either the family level or the package level. For example, starting a video recording by the camera is done by the void \textit{start()} method of the \textit{android.media.MediaRecorder}. In this case, the package name for this method is android, and the family name is media. The set of API calls used by each app, can be aggregated to a Control Flow Graph (CFG)~\cite{allen1970control}, a graph featured as a tree of API calls, based on package and family names (an example of a CFG is presented in Fig. \ref{fig:cfg}).

A thorough exploration of the dissassembled code and the CFG, along with the AndroidManifest xml file, allowed us to construct a set of novel attacks. A flaw in the manipulation of the APK may cause a crash in the repacking~\cite{berthome2012repackaging,jung2013repackaging} of the app, or when installing and running the app on the phone. While the manipulation in these cases succeeded in fooling the classifier, it had no attack impact.
\subsection{Androguard}
Several tools have been introduced over the years to fully examine an APK. In contrast to APKtool, Androguard~\cite{desnos2011androguard} is a popular open-source python based tool to dissect an APK. Androguard can be used both as a command line tool and an imported package for python scripts. A large portion of Android malware detection systems incorporate Androguard~\cite{alazab2020intelligent,AndroPyTool,ashawa2019host,badhani2019android,cam2019detect,chavan2019comparative,fatima2019android,gonzalez2020development,feng2019mobidroid,khoda2019selective,kumar2018famous,liang2019witness,ma2020droidetec,parkframework,qamar2019hybrid,qin2019msndroid,raghuraman2020static,sharmeen2020adaptive,sourav2019deep,wang2020android,watanabe2020study,wu2019malscan,xiao2020android}. When analyzing an APK, Androguard efficiently extracts features such as API calls, CFG graph, permissions, Activities, Package names, etc. 
Androguard analyzes both the Manifest file and the smali code files to extract specific features. It uses cross references (XREFs) to identify Classes, Methods, Fields and Strings. XREFs work in two directions: xref\_from and xref\_to. Xref\_to means that the current object is calling another object. xref\_from means that the current object is called by other object. These connections enable Androguard to visualized the XREFs as an directed graph (CFG). For the Manifest file, Androguard uses axml parser. For example, to get the requested permissions, Androguard parses the Manifest file and enumerates the permission names from the \textbf{uses-permission} tags. 
\section{Related Work}
\label{related}
In this section, we survey well known ML-based detection systems for Android malware. We discuss three main approaches. The first is a static analysis, which includes gathering string values from the Manifest file and the smali code files. The second approach follows the CFG of the application, which traces the connection between API calls used in the app. The third approach inspects the behavior of the app and gathers information on the OS's behavior during the run of the app, along with network usage, etc. 
 
In addition, we describe two evaluations of Android ML detection systems. We also depict several popular evasion attacks targeting the detection systems. We discuss three forms of attack. The first form engages in the art of camouflage, where the attacker tries to hide  strings and values of the app that may pinpoint malicious content using obfuscation and encryption. The second form adds noises to the app, e.g., non-invoked functions. The last form tries to change the flow of the app. It analyzes the previous flow of the app, and changes the code of a few function calls.

Next, we describe several Android malware permission based detection systems. Since this paper introduces several evasion attack vectors that focus on Android permissions, we discuss state of the art works that deal with these attributes.

Last, we examine the functionality and maliciousness tests of Android malware. To the best of our knowledge, these parameters have not been fully explored in previous works in the Android malware field. 

\subsection{Android malware ML-based detection systems}
\label{detection systems label}
One of the best-known Android malware detection systems is Drebin~\cite{arp2014drebin}. Drebin gathers syntactic features, and collects eight types of features from the APKs. Four features are extracted from the manifest file which include permissions requested from the user, software/hardware components residing in the app and intents that were run in the app. Four other features are gathered from the smali code, which include suspicious/restricted API calls, used permissions in the app's run and URL addresses. The DroidAPIMiner~\cite{aafer2013droidapiminer} is similar to Drebin, in that API calls and permissions are inspected. The authors performed a dataflow analysis to recover frequently used APIs and package names. 
They used multiple ML algorithms, where KNN dominated the other techniques with a 0.99 accuracy rate. A similar approach to DroidAPIMiner can be found in DroidMat~\cite{wu2012droidmat}, where other features from the AndroidManifest XML file were inspected, including the app components (such as intents and activities), aside from the app permissions. MalPat~\cite{tao2017malpat} analyzed the use of API calls in relation to permissions in the apps (based on the Pscout~\cite{au2012pscout} tool) to find malicious patterns. 
A recent example of syntactic feature analysis is Dejavu~\cite{salem2019don}, which merges static features analysis with code to compare benign against malicious apps. The Android malware detection systems using static analysis tap several features from the app. These features are easy to understand and deploy, and do not require a great deal of computing resources from the operator. However, since most of the static features can be mined efficiently, they can also be manipulated without consuming a lot of computing resources~\cite{demontis2017yes,maiorca2015stealth,meng2016mystique,rastogi2013droidchameleon}.

A different approach can be found in MaMaDroid~\cite{onwuzurike2019mamadroid}, which extracts features from the Control Flow Graph (CFG) of an application. MaMaDroid creates a tree of API calls based on package and family names. After abstracting the calls, the system analyzes the API call set performed by an app, to model its true flow. The authors built a statistical Markov chain~\cite{kemeny1976markov} model to represent the transitions between the API calls. Backes et al.~\cite{backes2016reliable} used a similar approach to detect malicious third party libraries in apps. 
A different perspective that used a similar approach to MaMadroid was presented by Sahs et al.~\cite{sahs2012machine}, who used Support Vector Machine (SVM)~\cite{suykens1999least,scholkopf2001learning} on permissions and CFG. The authors merged jumps and code blocks to reduce the CFG. They used several kernels on the outputs of the above, and labeled nodes in the CFG based only on the last instruction of the block it represents. Salem et al.~\cite{Loc_imp,salem2018idea} reported that API calls traces may be good indicators of malicious behavior. The authors injected anomalous API calls into benign API call traces. They calculated the likelihood of an API call sequence. A low value suggested a malicious app. The authors ran Droidutan~\cite{Droidutan} to get the trace of benign apps. Then, they injected several API calls as a block of operations. The authors used a Hidden Markov Model~\cite{eddy1996hidden,fine1998hierarchical,ghahramani1996factorial} to create a model from the benign apps. They achieved an accuracy rate ranging from 0.64-0.99. However, since they used blocks of codes, scattered malicious API calls remained unaddressed. Hou et al.~\cite{hou2018make} inspected API call sequences. The authors categorized the sequences by code blocks, package names and invoke types (from the smali~\cite{Smali_functions} code). These functions describe the similarity between API calls. The authors use HIN to fully analyze the connections between each category. They achieved a maximum accuracy rate of 98.6 while using the three categories. Monitoring the control flow/ API call set of the app seems like a better idea than the syntactic approach. Changing the flow of the app is more complicated, since the CFG may be complex. Furthermore, a change to the sequence of API calls of the app may damage the functionality of the malicious app, or its maliciousness. However, evasion attacks that manipulate the flow of the app such as as~\cite{chen2018android,ikram2019dadidroid,maiorca2015stealth,piao2016server,sun2014nativeguard} succeed in deceiving this kind of detection systems.

Several evaluations have been conducted on Android malware detection systems. Roy et al.~\cite{roy2015experimental} conducted a large survey on $\sim 1M$ apps and several Android malware detection systems (e.g., Drebin~\cite{arp2014drebin}) and malware datasets (like the Genome malware project ~\cite{zhou2012dissecting}). 
For verification, the authors used VT~\cite{total2012virustotal} (which we use in a similar manner). The results revealed a difference of 60\% in the true positive rate (TPR) between apps that were identified by at least 10 VT scanners and the apps that were identified by fewer than 10 scanners. 
The authors also examined Android malware detection systems to see whether the number of features could be reduced and still maintain a high detection rate. For example, the authors showed that Drebin's~\cite{arp2014drebin} feature set could be reduced to $\sim 2K$ and still preserves an accuracy rate of 99\%. 

Pendlebury et al.~\cite{pendlebury2018tesseract} conducted
a performance analysis on Android malware detection systems using Drebin,
MaMaDroid and Deep Neural Networks (DNN) algorithm~\cite{Grosse17}. The authors created
a metric to test the performance of these detection systems correctly. This metric implements ratios of old and new apps, and benign/malicious classes
in the training and test data. The ratio of 90/10 between benign and malicious apps
is based on an evaluation of the distribution of benign and malicious classes in
the real world. These proportions make the classifiers more robust to new malware. 
Note that the current study adopted some of the recommendations put forward in this paper. 


The third approach involves the use of several parameters such as CPU,  network usage, number of running processes, and battery levels along with other system features. This approach was introduced in Andromaly~\cite{shabtai2012andromaly}. The authors inspected two sets of apps: games and tools. The authors found that games could be differentiated more easily into benign and malicious because of the unique features of these apps. 
A similar approach was taken by
Shabtai et al. in~\cite{shabtai2014mobile,shabtai2010intrusion}. 
Behavioral detection methods are based on the behavior of the Android device while running the malicious and benign apps. These behaviors may be device-dependent. Therefore, a dataset based on behaviors may suffer from high diversity. In addition, applying these methods as a host-based method as was done in~\cite{shabtai2014mobile} without a global dataset will fail when a new device is used.

Nissim et al.~\cite{nissim2016aldroid} used Active Learning (AL) methods and a Support Vector Machine (SVM) classification algorithm to detect Android malware using static analysis. The authors tried to identify malicious signatures using AL along with
judgments by a human cyber expert. 
Although a human interaction enhances the detection 
rate, it interferes with the autonomy of the detection system and its automatic process.

For more information of Android malware ML detection systems, see~\cite{bakour2019android,kouliaridis2020survey}.

\subsection{Evasion attacks on ML-based detection systems} Evasion attacks against ML-based detection systems can take multiple courses. One course of action is to camouflage specific features of the app. One well-known example of camouflage is the use of obfuscation or encryption. Demontis et al.~\cite{demontis2017yes} used obfuscation of suspicious string values, APIs, and packages. Another example of camouflage is packing an app inside another app. DaDidroid~\cite{ikram2019dadidroid} used a similar approach as found in Demontis et al.~\cite{demontis2017yes} by adding packing to the obfuscation methods. The authors wrapped the original Android malware inside another app.

Reflection, which allows a program to change its behavior at runtime, is also a classic evasion method. Rastogi et al.~\cite{rastogi2013droidchameleon} presented an attack which uses the Demontis et al.~\cite{demontis2017yes} approach along with reflection. 

A typical approach to evasion attacks on ML-based detection systems includes adding noise to the app, thus misleading the classifier's assignment of benign and malicious app. 
An example of the use of this approach can be found in Android HIV~\cite{chen2018android} where the authors implemented non-invoked dangerous functions against Drebin and a function injection against MaMaDroid. Another option is to use Generative Adversarial Networks (GAN) to add noises to the classification process.  Shahpasand et al. ~\cite{shahpasand2019adversarial} used GAN to add features from the Drebin log file to a malicious app sample.

Changing the app flow is another approach. Here a detection system that is based on analyzing the app flow, like MaMaDroid, fails to detect the malicious app~\cite{chen2018android,ikram2019dadidroid}. 
Stub function\footnote{A stub function is a function that basically does not do anything, but may change the flow of an app.} can also be used to break the app flow ~\cite{chen2018android,piao2016server,sun2014nativeguard}, resulting in an ML misclassification of a malicious app. 

Our evasion attacks use obfuscation methods to conceal the appearance of API calls. In addition, we use reflection on the obfuscated API calls to implement a more complex adversary. Also, we enhance our attacks with special focus on the Manifest file, thus implementing new attack methods such as Manifest Pockets (which we will explain later) and other Mainfest based attacks. We do not incorporate noise or change of the app flow in our evasion attacks.  

\subsection{Permission-based detection systems}
System permissions are one of the core components of every operating system. Since Android is an OS, it incorporates system permissions requests from the user to get system resources as a uses-permission tag in the Manifest file containing a constant string name of a permission. A system resource that is used by an app but is not requested from the user will cause a failure. Therefore, correct permission requests are a vital component for an APK. Also, specific combination of permission requests may pinpoint malicious activity. For example, a combination of  the Internet and Phone state permission  requests may be a trace of stealing phone numbers and sending them to a third party site. 
For these reasons, the permission requests are considered as straightforward features for Android malware detection systems. In this section we will discuss several Android malware permission based detection systems.

Kirin~\cite{enck2009lightweight} was one of the first known Android malware permission based detection systems. This rule based security system for Android inspects the use of Android permissions of an app against a list of rules at install time. The authors evaluated their system against 311 apps from the Android market, where 10 apps raised alerts. Five of the suspicious apps asserted a dangerous configuration of permissions,
but were used within reasonable functional needs based on
application descriptions. The other 5 apps were categorized by the authors as potentially malicious and needing care when installed. Therefore, they concluded that their technique requires user involvement for approximately
1.6\% of the applications. Unfortunately, as the rules of Kirin are contant and specific for a small number of permission requests, new and updated malicious apps can easily evade this kind of detection system.  

Another approach in permission-based detection systems is the relations between permission requests. Rassameeroj et al.~\cite{rassameeroj2011various}  produced a permission adjacency matrix for every two permissions found in an app. These were converted to a graph where the nodes were permissions, and the weights on the edges represented the frequency of the corresponding
permission concurrence. The authors also calculated the apps' adjacency matrix. Each app was represented by a binary vector where each element represented the request of a specific permission. Based on the Euclidean distance between vectors as similarity, a graph was built, where the nodes were apps and the weights on the edges represented the similarities between apps. 
Using Kirin's~\cite{enck2009lightweight} dangerous permissions detection, they detected APKs could contain malicious operations. Therefore, their detection system suffers from the same exploits as Kirin. 

A more recent study on permissions in~\cite{arora2019permpair} described correlations between permission pairs. Here, detection is based on graphs of relations between permissions in malicious and benign datasets. In the train phase, each permission was converted to a node. For each sample, an edge with weight 1 was created between pairs of permissions included in the app. If the edge existed, its weight was increased by 1. In the testing phase, the sum of weights was calculated in both graphs. The label for the sample was determined by the maximum value between the sum of permission pairs in the malicious graph and the sum of permission pairs in the benign graph. As their study was static, they could not detect updating apps at runtime. The authors noted that apps with one or two permissions were misclassified. 

Droid Detective~\cite{liang2014permission} is a detection system based on Android permission combinations. This system runs from k=1 to 6 to find combinations of permission sets that can be found on malicious and benign apps. Then, a selection algorithm is run on the results. The selected sets are used as security rules. 
They found that the larger the value of k, the more accurate
the detection is. Almost all the permission combinations that they found to be malicious included a SMS affiliated permission. 

Droid Permission Miner~\cite{aswini2014droid} studied the permissions in the intersection of malicious and benign apps, the union of both, and in each class that was not in the other. The authors used Bi-Normal separation~\cite{forman2003extensive,forman2006bns} (BNS) and Mutual Information~\cite{battiti1994using} for their feature selection. Using BNS, they produced the top 30, 25, 20, 18 and 15 features and the bottom  30, 25, 20, 18 and 15 features as options for the classification algorithms. Their classification algorithms were Naïve Bayes, AdaBoost M1 with J48 as the base classifier, J48, Random forest and IBK-5. They found that the intersection group between the malicious and benign apps was the best identification group. 


The use of maliciousness score is explored in the permission-based detection systems. In APK Auditor~\cite{talha2015apk}, the authors presented a client-server based system that calculates a malware rank based on the app's requested permissions. A permission malware score (PMS) was calculated for each
permission using its existence in all malware apps. Then, an Application's Malware
Score (AMS) was calculated by summing all its permissions' PMSs. The authors used logistic regression~\cite{hosmer2013applied} to measure the malware threshold limit. A similar approach to the permissions risk score was introduced in \cite{peng2012using}. Forensic Analysis of Mobile devices Using Scoring
of application permissions (FAMOUS)~\cite{kumar2018famous}
introduced permission scoring in APKs. The authors used the scoring of permission requests to detect malicious and benign apps. They counted the occurrences of permission requests in both benign apps (PuB) and malicious apps (PuM). The authors divided both counters in the correlative group size (B for benign, M for malicious), and created BSP and MSP. An Effective Maliciousness Score of Permission (EMSP) was calculated as the final score or a permission: EMSP = MSP - BSP. They defined the Maliciousness score by accumulating the EMSPs of the permissions each app requested. They calculated the EMSPs and MS score of each app as features of ML algorithms they have implemented. They tested RF, SVM, KNN, NB, and CART, where RF dominated all other algorithms. 
Tree classifiers are one of the most popular machine learning algorithms. Several Android malware detection systems are solely based on these classifiers. Aung et al.~\cite{aung2013permission} proposed permission-based Android malware detection systems using tree classifiers. They used K-means~\cite{wagstaff2001constrained,krishna1999genetic} to cluster the data. Then, they used decision trees on each cluster. The decision tree classifiers were - j48~\cite{bhargava2013decision,ruggieri2002efficient}, RF, CART~\cite{denison1998bayesian,steinberg2009cart}. The most promising algorithm was RF. 
Droiddet~\cite{zhu2018droiddet,droiddet-implementation} focused on permissions, permission-rate, monitoring system events and some sensitive APIs. They introduced the PERMISSION-RATE formula, which is the number of permissions an app requested divided by the size of the smali code files of the apps. They posited that the PERMISSION-RATE of the malware should ordinarily be higher than the benign rate. An app that provides various functions means that it will require a large of number of permissions. They used Rotation Forests~\cite{rodriguez2006rotation,kuncheva2007experimental} as their classifier\footnote{Rotation Forest Classifier is an ensemble classifier where each base classifier is trained on the entire dataset.}. The rotation forest adopted principal component analysis (PCA)~\cite{wold1987principal} to handle the feature subset randomly extracted for each base classifier to intensify the diversity. To encourage the diversity, bootstrap samples were employed as the training set for the base classifiers, as in bagging. 

Finding the exact feature set is another course in malware detection. SIGPID~\cite{sun2016sigpid} tried to find the minimum subset of permissions for Android malware detection that maintained similar values in precision, recall, accuracy and F-score as using the full permission set. The authors evaluated the permissions that were frequently requested by benign apps and malicious apps. In addition, permissions that were paired in most apps were represented by only one of them. Their final permission set size was 22. 
ESID~\cite{jayanthienhanced} is an enhanced SIGPID system. The first stage in ESID is feature selection, where a specific permission with a frequency lower than c is eliminated both in the malicious and benign data. The authors reduced the features set to 38 - 47 features. 

Overall, the mutual hypothesis of these permission-based detection systems is that permissions must be declared by the uses-permission tags in the manifest files. There is no other option available. The detection systems must acquire these permission requests to work. However, as we explore in this study, this assumption is false. There is another option to request permissions from a user. We sample several permission-based detection systems in this study to prove our point.   


\subsection{Functionality and maliciousness tests}
The main goal of evasion attacks is to avoid detection by malware detection systems. However functionality and maliciousness tests can be conducted to assess their impact. These aspects were explored in
Repackman~\cite{salem2018repackman}, a tool that automates the repackaging attack. The authors studied 100 apps from Google play, and successfully repackaged 86\% of the apps with random payloads that maintained the functionality of the previous apps. They performed 2 tests on their repackaging attack, all using Droidutan~\cite{Droidutan}. For the functionality test (which they termed \textbf{feasibility}), they used random UI actions. For maliciousness, which they termed \textbf{reliability}, they measured the apps that successfully executed their payload via recorded screenshots. 
To the best of our knowledge, these functionality and malicious content tests have not been discussed in previous research in the field of Android malware evasion attacks.

Our functional app test is different in that we test the initial state of the app when it is started in an emulator. We do not interact with the app itself, because we evaluate its basic functionality. We assess the identification rate of VT scanners as an indication of maliciousness activity. Recent studies have began examining the strength of specific scanners against evasion attacks. This is adopted here. 
\section{Metrics and Evaluations} 
\label{eval}
This study implemented various metrics and evaluations. They were used to define the application dataset, assess the effectiveness of the evasion attacks against the detection systems, and formulate insights. 
First, we describe the dataset and its verification tool. We discuss dataset biases, and how we tackled them. We discuss functionality and maliciousness tests for Android evasion attacks. We show that if an app crashes on the functionality test, it will not cause any harm to the user who runs it.  For the maliciousness test, we compared the original Android malware to the manipulated Android malware to determine whether their reactions were similar. We tested original malicious apps and manipulated apps as regards identification by VT malware detection scanners.
The evaluations and metrics are:
\begin{itemize}
\item \textbf{Data collection:} 
    \begin{itemize}
        \item \textbf{Benign apps:} We used apps from the AndroZoo dataset~\cite{Allix:2016:ACM:2901739.2903508} ($\sim74K$)  chosen from GooglePlay \cite{GooglePlay} market, and CICAndMal2017's~\cite{lashkari2018toward} dataset ($\sim0.1K$). \\
        \item \textbf{Malicious apps:} We used the malicious apps from the Drebin dataset\\~\cite{arp2014drebin}($\sim5.56K$), CICAndMal2017~\cite{lashkari2018toward} ($\sim0.4K$), AMD~\cite{wei2017deep} ($\sim24.5K$) and StormDroid~\cite{chen2016stormdroid} ($\sim1.6K$) datasets.\\
        \item \textbf{Verification:} We used VirusTotal (VT)~\cite{total2012virustotal} to verify that our apps were correctly labeled (i.e., malicious or benign). We define benign apps as apps that are not marked as malicious by fewer than 3 scanners. Malicious apps are apps that are identified as malicious by at least 2 to 4~\cite{arp2014drebin,pendlebury2018tesseract} scanners (out of 64 available scanners). We used malicious apps from the above datasets that were identified by at least 3 scanners.\\ 
    \end{itemize}
\item \textbf{Eliminating dataset biases:}  
Android malware detection systems suffer from temporal and spatial biases~\cite{pendlebury2018tesseract}. Spatial bias is defined as unrealistic assumptions about the
ratio of benign to malicious apps in the data. Temporal bias is defined as temporally inconsistent
evaluations that integrate future knowledge about the test data into the training data.
 
To avoid these dataset biases, we followed the suggested properties in~\cite{pendlebury2018tesseract}. For example, 90\% of our dataset was composed of benign APKs, and the remaining 10\% were malicious, similar to the distribution of global benign and Android malware apps~\cite{Go_sec_2017,lindorfer2014andradar}, thus accounting for the spatial bias. 
As for the temporal bias, we trained the classifiers with apps whose timestamp was set prior to the test data.\\
\item \textbf{Robustness evaluation:} To evaluate robustness, we computed the proportion of malicious instances for which the classifier was evaded; this is our metric of \emph{evasion robustness}, with respect to the robustness of  the detection system (similar to the analysis provided in~\cite{tong2017framework}). Thus, an evasion robustness of 0\% means that the classifier was successfully evaded every time, whereas an evasion robustness of 100\% means that the evasion failed in every instance. We compared the the evasion robustness of each evasion attack to the original malicious APKs detection rates of each detection system. \\
\item \textbf{Functionality:}\label{func_metric} We evaluated the functionality of each manipulated app. In an evasion attack, the attacker tries to hide the malicious activity from the detection system. Constructing an evasion attack with existing Android malware involves changing the APK, which is a sensitive job. For example, a line that describes invoking a method: \textbf{"invoke-virtual\{v0, p2, v1\},Ljava/lang/Class;-$>$getDeclaredMethod(Ljava/lang/String;\newline[Ljava/lang/Class;)"}. This line assumes that the first argument stores a class object (v0), a string is stored in the second argument (p2), and an array of class objects is stored in the last argument (v1). If one of the arguments contains primitives or different objects than stated, the app crashes. This is defined as to a situation in which an error occurs during an app's run which lead to the app's process kill. Automation of feature changes may result in this kind of error (or other errors), resulting in a crash. If the crash occurs in early stages of the run, the malicious content will probably not damage the user at all. Therefore, evaluation of the app's functionality is vital when generating an APK evasion attack. The functionality test has two steps: the app is installed and run on an Android emulator~\cite{emulator}. If the app crashes, it is concluded that it is not functional. 
A manipulated app that passes this test is declared a functional app.\\
\item \textbf{Malicious Activity/Maliciousness:}\label{malic_metric} A maliciousness check is an extension of a functionality check. Whereas a manipulated app that passes the previous check can be considered a threat to the user, it does not guarantee that the former malicious content will run similarly to the former Android malware. Some of the activities the malware runs may fail or change drastically as a result of the manipulation of the Android malware. Therefore, a maliciousness test of the app is needed. The maliciousness check measures the similarity between the previous Android malware and their manipulated counterparts. These tests took place in the following steps. First, we scanned the original malicious app using VT. Then, we scanned the manipulated app in the same way. If the number of scanners that detect the manipulated app was less or equal to the number of scanners identifying the original malicious app, this app passed our maliciousness test. 
\item \textbf{Permissions inspection:}
     \label{perms_ins}
We analyzed the use of permissions in the dataset. This was done by splitting the inspection into two \textbf{groups}: benign apps and malicious apps to better understand how the permissions requests for each group differed. We used Android's \textbf{protection levels} (see in Section \ref{Intro}) to map the behavior of each group. We inspected two properties of each group:
\begin{enumerate}
    \item \textbf{General protection level use:} We extracted the requested permissions from each AndroidManifest.xml file of each app. Then we mapped the permissions to their protection levels and accumulate the use of each protection level from the application to its group. 
    \item \textbf{Dominant protection level:}  The same process as in the general case, but this only accumulated the dominant protection level of each app. In other words, the goal was to find the most frequent protection level used in each app and assign it to the relevant group.\\
    \end{enumerate} 
    
    \item \textbf{Permission Families:}\label{perm_family} We  analyzed the statistics of permissions use in the benign and malicious groups and to do so created \textit{permission families}. Each family contained at least one permission. The families were created in a recursive approach. This consisted of inspecting each permission separately to obtain their distribution in the group. Then, for each one of them, a different permission was added recursively and the distribution of the permission pair was inspected. This process was then extended to permission triplets, permission quadruplets, etc. which we term permission families. The rationale for creating these families is to understand the behavior of permissions in each group. An attacker that acquires this information can create an evasion attack that mimics the use of permissions such as a benign app.

\end{itemize}
\section{Attacker models}
\label{model}
In this section, we describe three attacker models. The first two models depict an attacker that has black-box~\cite{madry2017towards,papernot2017practical,shahpasand2019adversarial} access to the trained classifier. The third model is a zero knowledge attacker. We describe the initial evaluation of the trained classifier, which is needed to assess the baseline for the evasion attacks.
Then we describe the two types of data used in our attacker models: Drebin reports and Permission families statistics. This is followed by a description of the attacker models: Model Access, Data Access and Zero Knowledge. To enhance the readability of Sections \ref{model} and \ref{attacks}, an abbreviation table is provided in Table~\ref{abrev}.

\begin{table}[!h]
	\centering
	\caption{Abbreviations for the Attacker models and Evasion attacks}
	\label{abrev}
	\begin{tabular}{l|c}
	    \hline
		 MA & Model Access (attacker model). \\
		DA & Data Access (attacker model). \\
		ZK & Zero Knowledge (attacker model). \\
		MB & Manifest Based (evasion attack). \\
		SB & Smali Based (evasion attack). \\
	\end{tabular}
\end{table}

\subsection{Initialization phase}
\label{initialization_phase}
\begin{figure}[ht]
	\centering
	\includegraphics[width=0.7\columnwidth]{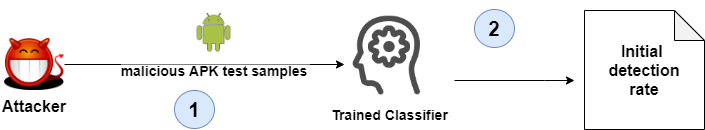}
	\caption{The initial step of the attack models. The attacker sends malicious APK test samples to the trained classifier (1). Then, the classifier produces a set of labels of the apks. The attacker  accumulates these labels to formulates the detection rate as an initial detection rate of the malicious APKs (2), which we term the \textbf{Initial detection report}.}
	\label{fig:init_attack_model}
\end{figure}

 Each attacker model started with an initial evaluation of the classifier detection rate, as depicted in Fig.~\ref{fig:init_attack_model}. The following steps were carried in this phase: \begin{enumerate}
     \item The attacker sends malicious APK test samples to a trained classifier. The trained classifier was trained on our benign and malicious APK dataset which was described in Section~\ref{eval}. The classifier returns a label for each APK. 
     \item The adversary accumulates these outputs to come up with the detection rate, which we denote the \textbf{initial detection rate}. 
 \end{enumerate}
\subsection{Drebin reports}  
\label{drebin_reports}
Drebin is a lightweight method for the detection of Android malware. In this section, we describe the report Drebin produces on APKs. For a full description of the Drebin  classifier, see~\cite{arp2014drebin} (implementation is available at~\cite{Drebin_implementation}). 

During its run, Drebin creates the following observations on each APK sample:
\begin{enumerate}
    \item \textbf{Component lists:}
The observations included are \textit{IntentActionList}, \textit{ServiceList}, \textit{ActivityList}, and \textit{BroadcastReceiverList}. These observations are obtained from the manifest file by following the corresponding components tags; namely, Intent, Service, Activity, Receiver.

\item \textbf{Requested Permissions:}
These observations describe the permissions that were requested from the user. Drebin maps them according to the \textbf{Uses-Permission} tags from the manifest file.
\item \textbf{Suspicious APIs:} These observations in the Drebin report relate to a constant number of APIs defined in the system. Examples of suspicious API calls are \textit{setWifiEnabled()} and \textit{sendTextMessage()}.

\item \textbf{Restricted APIs and Used Permissions:} The classifier includes a dictionary called \textit{SmallCasePScoutPermApiDict.json} that maps API calls to permissions. Drebin looks for these calls to understand which permissions are used by the app. Next, it lists the permissions that correspond to permissions that were requested in the manifest as \textbf{Used Permissions}. Finally, it adds the API calls that were mapped to permissions that were not mentioned in the manifest file as \textbf{Restricted API} calls.\\
\item \textbf{URLDomains:} The classifier lists the URLs mentioned in the smali code files. 
\end{enumerate}

\begin{figure}[!]
	\centering
	\includegraphics[width=0.7\columnwidth]{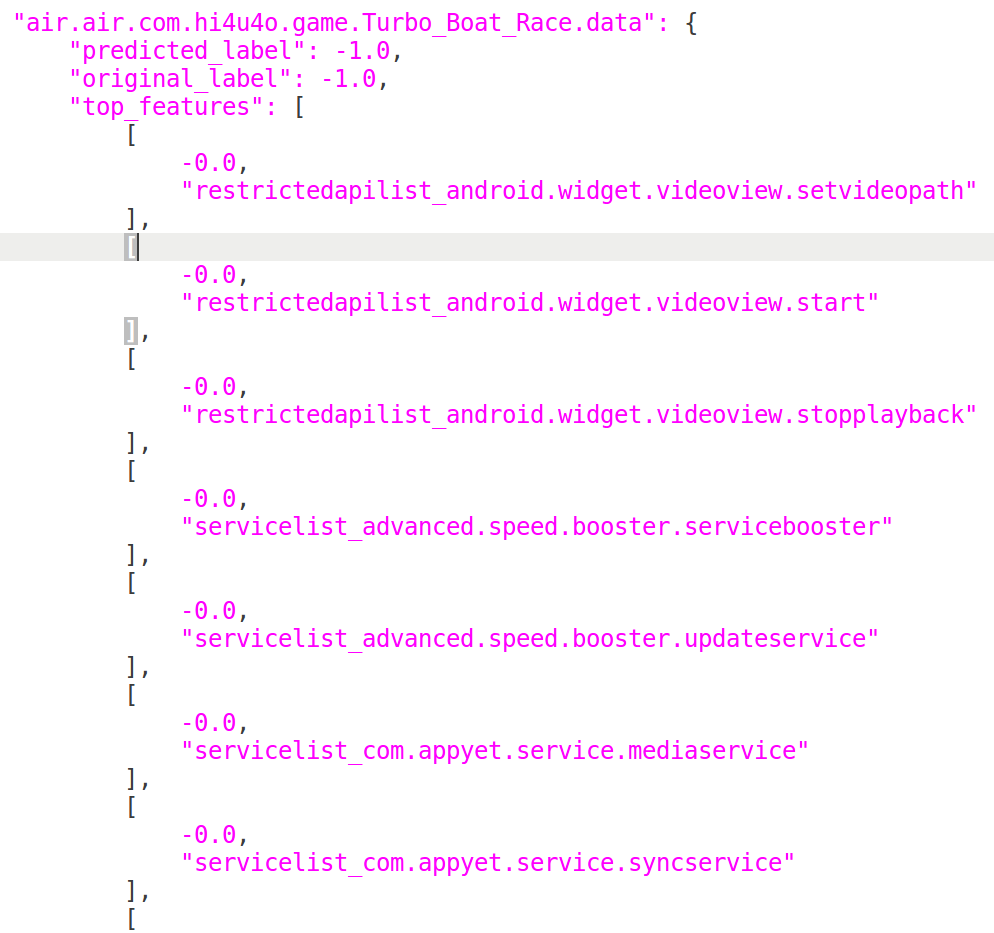}
	\caption{Drebin report of an APK. This report is associated with a benign app, since the original label is -1, which stands for benign. As can be seen, the weights for each observation are 0. Therefore, the predicted label for the app is -1. }
	\label{fig:rep}
\end{figure}

Overall, the steps in Drebin's report creation are as follows:
\begin{enumerate}
    \item Drebin saves the original labels of each app. 
    \item During the training process, Drebin inspects each training sample (APK file). It assigns each observation a corresponding weight $W_s$ and sums these weights.
    \item Using SVM, Drebin computes the maximum weight that corresponds to a benign app as $t$.
    \item In the test phase, Drebin scans the test samples, labels each feature in each app with a weight according to its training phase. It saves the weights and labels in \textit{Explanations.json} (see Fig. \ref{fig:rep} for an example).
    \item In the test phase, if $\Sigma W_s > t$, Drebin predicts a label of 1 (malicious). Otherwise, it predicts the label of -1. It saves the weights and labels in \textit{Explanations.json} (see Fig. \ref{fig:rep} for an example), which saves the \textbf{Drebin reports}.
    
\end{enumerate}    

An attacker that has access to Drebin's reports is able to scan the output of the trained Drebin classifier on samples it obtained and sent to Drebin. The adversary has only a black-box access to Drebin. In addition, the attacker focus solely on the observations with $W_s>0$. Since an observation associated with $W_s=0$ is not considered important in Drebin, it is not included in the evasion attack process.
\subsection{Permission family statistics}
The permission families, as described in Section~\ref{eval}, classify the distribution of permission requests into benign and malicious apps in the data. An attacker that has access to these statistics understands how to resemble a benign app. For example, if a top permission family is [\,X,Y,Z]\, and a malicious app requests permission  [\,A,X,B,Y,Z,C]\,, the attacker will conceal permission requests [\,A,B,C]\, to mimic the benign app.
\subsection{MA attacker model}
\label{ma_attacker}
The first attacker model is called \textbf{M}odel \textbf{A}ccess. This model has access to Drebin's reports (and therefore, to the APKs' classifications). Drebin maps significant syntactic features from each APK in a report (described in detail in Section~\ref{drebin_reports}). Therefore, its output reports focuses the attacker on the important features it needs to conceal. The attacker does not need the data itself (unlike in to Section~\ref{da_attacker}). It is guided by Drebin. The attacker follows the following steps:
\begin{enumerate}
    \item The attacker uses the malicious APK test samples as input to Drebin.
    \item Drebin outputs its reports.
    \item The attacker uses Drebin's report as part of the manipulation process. It focuses on the significant observations.
    \item The manipulated APKs are sent as input to the trained classifier.
    \item The classifier produces the labels from the APKs. The attacker accumulates them and generates the  final detection rate.
\end{enumerate} Finally, the attacker compares the final detection rate with the \textbf{Initial detection rate} from the initialization phase (Section ~\ref{initialization_phase}).
\begin{figure}[ht]
	\centering
	\includegraphics[width=0.7\columnwidth]{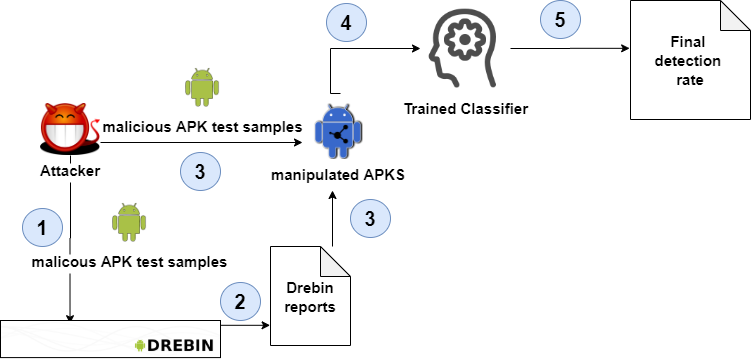}
	\caption{Model Access attacker model - The attacker runs a Drebin trained classifier with the malicious APK test samples (1). Drebin produces a report for each APK (2). Then, the attacker analyzes the reports to decide which features it needs to manipulate to create the manipulated APKs (3). The manipulated APKs are used as input to the trained classifier (4). Then, the classifier produces the set of labels of the APKs. The attacker accumulates them to create a detection rate of the manipulated APKs (5). The attacker compares this detection rate to the \textbf{Initial detection rate} from the initialization phase (Section ~\ref{initialization_phase}).}
	\label{fig:ma_attack_model}
\end{figure}
\subsection{DA attacker model}
\label{da_attacker}
The second attacker model is called \textbf{D}ata \textbf{A}ccess. This model has access to the dataset used to train the classifier. 
As discussed in Section~\ref{eval}, permission families can help understanding the deep connections between permissions in benign and malicious apps. The attacker proceeds as follows: 
\begin{enumerate}
    \item The adversary analyzes the train apps to get the permissions families.
    \item The attacker uses insights on the permission families from the benign train data to efficiently manipulates the malicious APK test samples.
    \item The manipulated APKs are sent to the trained classifier.
    \item The classifier produces the set of labels for the manipulated APKs. The adversary generates the final detection rate from accumulation of the labels. 
\end{enumerate} Finally, the attacker compares the final detection rate to the \textbf{Initial detection rate} from the initialization phase (Section ~\ref{initialization_phase}). 
\begin{figure}[!]
	\centering
	\includegraphics[width=0.7\columnwidth]{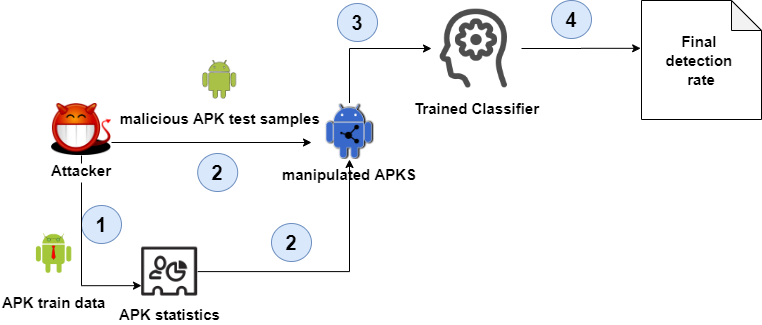}
	\caption{Data Access attacker model - The attacker gains insights on the permission families from the benign train data (1). It uses this information to efficiently manipulates the malicious APK test samples (2). The manipulated APKs are sent to the trained classifier (3). Finally, the classifier produces labels for the manipulated APKs. The adversary accumulates these labels and generates the final detection rate (4). The attacker compares this detection rate to the \textbf{Initial detection rate} from the initialization phase (Section ~\ref{initialization_phase}).}
	\label{fig:da_attack_model}
\end{figure}
\subsection{ZK attacker model}
\label{zk_attacker}
The third attacker model is called \textbf{Z}ero \textbf{K}nowledge. This model has access only to the trained classifier as a blackbox, with no access to any external data. The adversary follows the following steps:
\begin{enumerate}
    \item The attacker blindly manipulates the malicious APK test samples.
    \item  The manipulated APKs are used as input for the trained classifier.
    \item The classifier produces a set of labels for the manipulated APKs. The attacker accumulates the labels and creates a final detection rate.
\end{enumerate}
   Finally, the attacker compares the final detection rate to the \textbf{Initial detection rate} from the initialization phase (Section ~\ref{initialization_phase}). 
\begin{figure}[ht!]
	\centering
	\includegraphics[width=0.7\columnwidth]{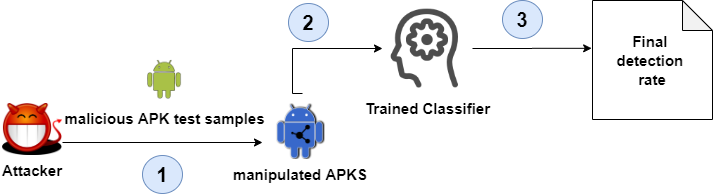}
	\caption{Zero Knowledge attacker model - The attacker blindly manipulates the malicious APK test samples (1). The manipulated APKs are used as an input for the trained classifier (2). Finally, the classifier produces a set of labels for the APKs. The adversary generates the final detection rate (3) from the accumulation of the labels. The attacker compares this detection rate to the \textbf{Initial detection rate} from the initialization phase (Section ~\ref{initialization_phase}).}
	\label{fig:zk_attack_model}
\end{figure}
\section{Evasion attacks}
\label{attacks}
Based on the attacker model, an evasion attack should be engineered that transfers the embedded knowledge of the model to a manipulated APK that will be classified as benign. In this section we suggest a set of evasion attacks. Each attack is characterized using a tuple, which we term Attack tuple (Section~\ref{attack_tuple}). Then, we describe a general attack template in Section~\ref{general_attack_template}, that is common to all the attacks. Then we describe each attack.
\subsection{Attack tuple}
\label{attack_tuple}
Each attack in this paper integrates different parameters.  Therefore, we define the \textbf{Attack Tuple}
(AM,S,M), where: 
\begin{enumerate}
    \item 
     AM represents the attacker model: Model Access attacker (as described in Section~\ref{ma_attacker}), Data access attacker (as described in Section~\ref{da_attacker}), and Zero-knowledge attacker (as described in Section~\ref{zk_attacker})
    \item S stands for access to the Smali code files. The attacker may or may not have access to the smali code files of the APKs it tries to manipulate. In other words, S defines whether the attacker manipulates the smali code files. We use a binary bit, where 1 means access granted and 0 means no access to the smali files.
    \item M refers to access to the Manifest file. The adversary may or may not have access to the Manifest file it wants to manipulate. We use a binary bit, where 1 indicates access granted and 0 indicates no access to the Manifest file.
\end{enumerate}
\subsection{General Attack template}
\label{general_attack_template}
In this section, we describe a general attack template. Each attack vector in our study manipulates other features. However, a basic template is implemented in each which follows Algo. \ref{alg:att_temp}.  The template consists of the following steps:
\begin{enumerate}
    \item The algorithm's input is an APK and additional data if present (Drebin report/permission families list).
    \item Depackage the APK to the Manifest file, smali code files and other subordinate files (line~\ref{lst:line:dpk}).
    \item Run the attack vector using the Manifest file, smali code files, other subordinate files and the additional data (line~\ref{lst:line:attack}).
    \item Repackage the APK (line~\ref{lst:line:repack}), and return it as an output (line~\ref{lst:line:end}).
\end{enumerate}
\begin{algorithm}[!]
\caption{General attack template }\label{alg:att_temp}
\begin{algorithmic}[1]
\Procedure{General Attack}{$APK,[Add\_data]$}
    \State $Manifest,Smali... \gets depackage(APK)$ \label{lst:line:dpk}
    \State $Manifest,Smali... \gets Attack\_Vector(Manifest,Smali...,[Add\_data])$ \label{lst:line:attack}  
\State $APK \gets Repackage(Manifest,Smali...)$\label{lst:line:repack}
\State \textbf{return} $APK$\label{lst:line:end}
\EndProcedure
\end{algorithmic}
\end{algorithm}

\subsection{Manifest Based attack vectors}
\label{mani_meth}
 For these attack vectors, the attacker exploits a weak spot in the classifier; namely, the fact that it does not check for any manipulated content in the Manifest file. The following attack vectors manipulated parts of the Manifest file, mostly the permissions the app requests. Definitions are provided first. Then, we describe four attack vectors that focus on the Manifest file.
 
\begin{figure}[!h]
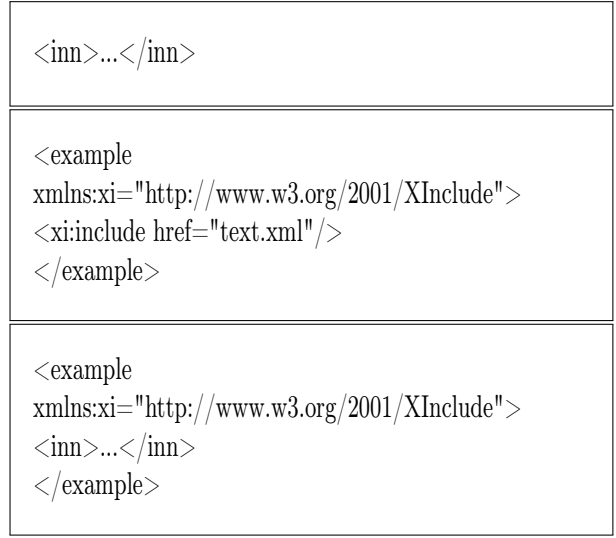

    \centering
    \adjustbox{minipage=8 cm, margin=1em,width=8cm,height=0.8cm,frame,center}
    {
    <inn>...</inn>}
    \adjustbox{minipage=8 cm, margin=1em,width=8cm,height=1.5cm,frame,center}
    {<example\\ xmlns:xi="http://www.w3.org/2001/XInclude">\\
  <xi:include href="text.xml"/>\\
</example>}
    \adjustbox{minipage=8 cm, margin=1em,width=8cm,height=1.5cm,frame,center}
    {\raggedright <example xmlns:xi="http://www.w3.org/2001/XInclude">\\
  <inn>...</inn>\\
</example>}
    \caption{Inclusion tag example. The first box is the contents of test.xml file. The second shows an example tag that includes a xi:include inner tag. In the last box, the included tag is replaced at runtime by the included test.xml file contents.}
    \label{fig:inclusion}
\end{figure}

 The native permission tag in the Manifest file is \textbf{Uses-Permission}. This tag specifies a system permission that the client has to grant the app. A newer version (on Android version 6.0+) of the uses-permission tag is \textbf{Uses-Permission-sdk-23}. From Android version 6.0+, \textbf{Permission Groups} were introduced. These groups place a number of sub-permissions requests together in the same permission request. For example, the READ\_SMS and WRITE\_SMS are included in a SMS permission. \textbf{XInclude} is an unfamiliar mechanism in XML, which merges XML documents by including one XML file in the other. An inclusion example is depicted in Fig. \ref{fig:inclusion}. The included XML content is not explicitly duplicated. Instead, the inclusion takes place during the parsing of the XML file and is written as an inclusion tag with a reference to a file where the content is placed. 
 The Android operating system needs the content of the Manifest file to run the app. When the classifier inspects the Manifest file it searches for specific features. It does not run a full parsing process on the Manifest file. Therefore, it ignores XInclude tags.
 One may assume that this attack results in a functional app since Android parses the Manifest file, including the merger of inclusion tags. However, we noticed that Android ignores these tags but the included content is not included at all, which makes the app non-functional. This phenomenon was observed in all XInclude uses.
The attack vectors included in the manifest based attack vectors are:
\begin{enumerate}
    \item \textbf{Manifest based attack type 1 (MA,0,1):} The attacker scans the report from Drebin (full details in Section~\ref{drebin_reports}) in parallel to the Manifest file. For the observations from the Requested Permission section, it  uses one of the following directions to conceal the requested permissions:
\begin{enumerate}
    \item Switch the uses-permission constant permission to its superior group constant. Most of the classifiers follow the rules of permission definitions from 2014. Therefore, they do not map these new permission definitions\footnote{We found that Android emulators with SDK 29 that correlates to Android 6.0+ ignore these permission requests as well. Therefore, the user was not asked to grant these permissions, and when such permission was needed by the app, the app crashed.}.
    \item \label{cuts}Cut  the permissions from the AndroidManifest.xml file. Then, create another XML file, say AndroidMan.xml and paste the permissions into AndroidMan.xml. Put an inclusion tag in place of the former permissions in the AndroidManifest.xml.
\end{enumerate}
For observations regarding components of the app (activities, receivers, intents),  the attacker moves the component tag from the Manifest file to a newly created XML file. Finally, it inserts a XInclude tag to replace the former component tag in the original file. The complete attack procedure  is depicted in Algo.~\ref{alg:mb1_algo}, which follows the following steps:
\begin{enumerate}
    \item The function’s input are the files depackaged from the APK and a list of observations produced by Drebin on this APK. 
    \item For each observation $s$ with $W_s$ > 0, run lines 3-10 (line~\ref{lst:line:loop_mb1}).
    \item Find the observation in the Manifest (line~\ref{lst:line:find_man_mb1}).
    \item If it is a component (line~\ref{lst:line:if_comp}), the chosen function is XInclude (line~\ref{lst:line:comp_con}).
        \item Otherwise, it is a permission (line~\ref{lst:line:if_perm}). Run \textbf{Random\_Func} to choose randomly between replacing the permission constant with its superior group constant to XInclude mechanism  (line~\ref{lst:line:perm_xinclude_mb1}).
    \item  Run the chosen function on observation $s$ and update  the manifest (line~\ref{lst:line:perm_con_mb1}).
    
    \item Return the manipulated manifest file and all the other depackaged files as output (line~\ref{lst:line:end_mb1}).
    
\end{enumerate}

\begin{algorithm}[!]
\caption{Manifest based attack type 1 function }\label{alg:mb1_algo}
\begin{algorithmic}[1]
\Procedure{MB1\_Attack}{$Manifest,Smali...,S$}
      \For{\texttt{$s \in S$
      s.t. $W_s>0$}} \label{lst:line:loop_mb1}
\State $s \gets find\_line(s,Manifest)$\label{lst:line:find_man_mb1}
    \If{\textit{s} is a component}\label{lst:line:if_comp}
    \State $f \gets XInclude$ \label{lst:line:comp_con}
    \Else \label{lst:line:if_perm}
        \State $f \gets Random\_Func(XInclude,Permission\_Group)$ \label{lst:line:perm_xinclude_mb1}
        \EndIf
    \State $Manifest \gets f(s,Manifest)$ \label{lst:line:perm_con_mb1}

\EndFor
\State \textbf{return} $Manifest,Smali...$\label{lst:line:end_mb1}
\EndProcedure
\end{algorithmic}
\end{algorithm}
\item \textbf{Manifest based attack type 2 (MA,0,1):} The attacker scans the report from Drebin (full details in Section~\ref{drebin}) in parallel to the Manifest file, and finds the Requested Permission section in the scan that is correlated to the Manifest file. Then, it switches the uses-permission tag of the permission requests to a uses-permission-sdk-23 tag. The complete attack procedure is shown in Algo.~\ref{alg:mb2_algo}, which follows the following steps:
\begin{algorithm}[!]
\caption{Manifest based attack type 2 function }\label{alg:mb2_algo}
\begin{algorithmic}[1]
\Procedure{MB2\_Attack}{$Manifest,Smali...,S$}
      \For{\texttt{$s \in S$
      s.t. $W_s>0$}} \label{lst:line:loop2}
\State $s \gets find\_line(s,Manifest)$\label{lst:line:find_man2}
    \State $Manifest \gets SDK\_23(s,Manifest)$ \label{lst:line:perm_con2}
\EndFor
\State \textbf{return} $Manifest,Smali...$\label{lst:line:end2}
\EndProcedure
\end{algorithmic}
\end{algorithm}

\begin{enumerate}
    \item The function’s input are the files depackaged from the APK and a list of observations produced by Drebin on this APK. 
    \item For each observation $s$ with $W_s$ > 0, run lines 4-5 (line~\ref{lst:line:loop2}).
    \item Find the observation in the Manifest (line~\ref{lst:line:find_man2}).
    \item \textbf{SDK\_23}: Change the uses-permission tag in $s$ to uses-permission-sdk-23 tag and save it to the Manifest file (line~\ref{lst:line:perm_con2}).
    
    \item Return the manipulated manifest file and all the other depackaged files as output (line~\ref{lst:line:end2}).
    
\end{enumerate}
\item \textbf{Manifest based attack type 3 (DA,0,1):} The attacker scans the permission family statistics in parallel to the Manifest file, finds the top 3 families from the benign apps dataset it wants to mimic. Then, it  correlates the Manifest file according to these families. For each app, it randomly picks a family to mimic. The attacker switches the uses-permission tags of all the permission tags that do not match a member of the family to a uses-permission-sdk-23 tags. The complete attack procedure is shown in Algo.~\ref{alg:mb3_algo}. The algorithm takes the following steps:
\begin{algorithm}[!]
\caption{Manifest based attack type 3 function }\label{alg:mb3_algo}
\begin{algorithmic}[1]
\Procedure{MB3\_Attack}{$Manifest,Smali...,F$}
    \State $f \gets Random\_Family(F)$ \label{lst:line:perm_family_mb3}
    
\For{\texttt{$p \in Manifest$
      }} \label{lst:line:loop_mb3}
        \If{$p \in f$}\label{lst:line:man_mb3}
        \State continue
        \Else
        \State $Manifest \gets SDK\_23(p,Manifest)$ \label{lst:line:perm_con_mb3}
        \EndIf
     
\EndFor
\State \textbf{return} $Manifest,Smali...$\label{lst:line:end3}
\EndProcedure
\end{algorithmic}
\end{algorithm}

\begin{enumerate}
    \item The function’s input are the files depackaged from the APK and the permission family statistics. 
    \item \textbf{Random\_Family}: choose randomly among the top 3 families in the benign app statistics (line~\ref{lst:line:perm_family_mb3}).
    \item For each permission $p$ in the manifest file, run lines 4-7 (line~\ref{lst:line:loop_mb3}).
    \item If $p$ is part of the family, continue (line~\ref{lst:line:man_mb3}).
    \item \textbf{SDK\_23}: Change the uses-permission tag in $p$ to uses-permission-sdk-23 tag and save it to the Manifest file (line~\ref{lst:line:perm_con_mb3}).
    
    \item Return the manipulated manifest file and all the other depackaged files as output (line~\ref{lst:line:end3}).
    
\end{enumerate}
\item \textbf{Manifest based attack type 4 (ZK,0,1):} The attacker blindly changes all uses-permission tags to uses-permission-sdk-23 tags. The complete attack procedure is shown in Algo.~\ref{alg:mb4_algo}. The algorithm implements the following steps:
\begin{algorithm}[!]
\caption{Manifest based attack type 4 function }\label{alg:mb4_algo}
\begin{algorithmic}[1]
\Procedure{MB4\_Attack}{$Manifest,Smali...$}

\For{\texttt{$p \in Manifest$
      }} \label{lst:line:loop_mb4}
        \State $Manifest \gets SDK\_23(p,Manifest)$ \label{lst:line:perm_con_mb4}
\EndFor
\State \textbf{return} $Manifest,Smali...$\label{lst:line:end4}
\EndProcedure
\end{algorithmic}
\end{algorithm}

\begin{enumerate}
    \item The function’s input are the files depackaged from the APK. 
    \item For each permission $p$ in the manifest file, run line 3 (line~\ref{lst:line:loop_mb4}).
    \item \textbf{SDK\_23}: Change the uses-permission tag in $p$ to uses-permission-sdk-23 tag and save it to the Manifest file (line~\ref{lst:line:perm_con_mb4}).
    
    \item Return the manipulated manifest file and all the other depackaged files as output (line~\ref{lst:line:end4}).
    
\end{enumerate}
\end{enumerate}

To sum up, MB1 uses XInclude and the superior  permission group name. The MB2, MB3 and MB4 attacks use uses-permission-sdk-23 tags. MB2 uses this tag by tapping Drebin reports. MB3 uses this tag by exploiting the permission families statistics. MB4 blindly conceals all permission requests by using the uses-permission-sdk-23 tag.
\subsection{Smali based attacks} 
\begin{enumerate}
    \item \textbf{Smali Based attack (MA,1,0):}
The attacker scans the Drebin's report (full details in Section~\ref{drebin}) along with the smali code files. It looks for Drebin observations related to smali code files, and specifically the \textit{Restricted APIs}, \textit{Suspicious APIs} and the \textit{Used permissions}. The attacker obfuscates any occurrence of the observation from the smali code files using several obfuscation methods. Note that each method is a prerequisite for the next method which is why we term each method a stage, where the input for the first stage is a string. This string can resemble a URL address or an API call. The attack vector stages revolve around encoding, reflection and manifest pockets:
\begin{enumerate}
    \item\textbf{First stage - Encoding:}
\label{enc_meth}
The attacker analyzes Drebin's observations (report) and looks for strings in the smali code that are included in the report; for example, a domain name, as shown in Fig.~\ref{fig:smali_encoding}. The attacker replaces the string  with its base64 encoding~\cite{josefsson2006base16}, and stores the encoded string in the variable of the former string (v5)\footnote{If the string represents an IP address, the attacker does not encode it simultaneously. It translates it into its integer representation and then encodes it. This is done because the IP address is basically a long integer. Therefore, encoding it as an address will produce an error in future decoding.}.
Next, the attacker runs a decode function (Landroid/util/Base64;->decode) to translate  the encoded string back to its former representation. Note that the decoded string is stored in the previous variable of the string (v5).

\begin{figure}[!h]
    \centering
    \adjustbox{minipage=7 cm, margin=1em,width=8cm,height=1.5cm,frame,center}
    {\raggedright \textbf{const-string v5, "http://abc.com"}\\
invoke-virtual \{p0, v6\}, Lwap/cash/DownloadActivity;\\->getSystemService(Ljava/lang/String;)\newline Ljava/lang/Object;}
    \adjustbox{minipage=8 cm, margin=1em,width=8cm,height=2.5cm,frame,center}
    {\textbf{const-string v5, "aHR0cDovL2FiYy5jb20="\\}
\textbf{const/4 p1, 0x0\\
invoke-static\{v5,p1\},Landroid/util/Base64;->decode(Ljava/lang/String;I)[B\\
move-result-object v5}\\
invoke-virtual\{p0,v6\}\newline ,Lwap/cash/DownloadActivity;\\->getSystemService(Ljava/lang/String;)\newline Ljava/lang/Object;}
    \caption{Encoding a string on smali code. The "http://abc.com" string in the upper box is replaced by its base64 translation. Then, a decode function is invoked on the encoded string. The decoded string is placed in the former variable of the string.}
    \label{fig:smali_encoding}
\end{figure}

\item\textbf{Second stage - Reflection:}
\label{ref_meth}
Reflection is a well-known method in Android malware~\cite{Fogla06,rodrigues2009robustness}. This stage is applicable solely for encoded strings that resembles an API call, whereas strings that resemble a URL proceed to the next stage. In this stage, the attacker creates a reflection call. The reflection call creates a general object with the target object from the API call. The next step is invoking the general object with the specific method name from the API call. An example of a reflection call that replaces an API call is depicted in Fig. \ref{fig:smali_reflection}. As mentioned earlier, the API call is encoded and then transferred to the reflection object. Our reflection object was implemented with a decoding method that decodes the encoded string to its former string so that it can be invoked.
\begin{figure}[!t]
    \centering
    \adjustbox{minipage=7 cm, margin=1em,width=8cm,height=1.5cm,frame,center}
    {const-string v5, "http://abc.com"  \\
\textbf{invoke-virtual \{p0,v6\},\newline Lwap/cash/DownloadActivity;->getSystemService(Ljava/lang/String;)\newline Ljava/lang/Object;}}
    \adjustbox{minipage=7 cm, margin=1em,width=8cm,height=3.5cm,frame,center}
    {\raggedright const-string v5, "http://abc.com"\\
...\\
\textbf{invoke-virtual \{v0, p2, v1\}, Ljava/lang/Class;->getDeclaredMethod(Ljava/lang/String;\newline[Ljava/lang/Class;)\newline Ljava/lang/reflect/Method;} \\
move-result-object p2 \\
... \\
new-array v0, v2, [Ljava/lang/Object; \\
\textbf{invoke-virtual\{p2,p0,v6\},\newline Ljava/lang/reflect/Method;->invoke(Ljava/lang/Object;\newline [Ljava/lang/Object;)\\Ljava/lang/Object;}}
    \caption{Reflection method on smali code. The getSystemService in the upper box is replaced with a general object based on DownloadActivity and the function. Then, it invokes the function.}
    \label{fig:smali_reflection}
\end{figure}

\item\textbf{Third stage - Manifest Pockets:}
\label{poc_meth}
In the third stage, the attacker uses the encoded string's content to create a meta-data tag in the Manifest. The tag has an arbitrary name, and the value of the string. Next, the attacker replaces the original smali code with a code that acquires the string from the meta-data tag (see Fig. \ref{fig:manifest pockets} for an example).
\begin{figure}[!]
    \centering
    \adjustbox{minipage=7 cm, margin=1em,width=8cm,height=1.5cm,frame,center}
    {const-string v4, \textbf{"100.50.43.22"}\\
const/16 v1, 0x22b8\\
invoke-direct\{v2,v4,v1\}\newline LJava/net/InetSocketAddress;-><init>(LJava/lang/String;I)V}
    \adjustbox{minipage=7 cm, margin=1em,width=8cm,height=2.6cm,frame,center}
    {\raggedright <?xml version="1.0" encoding="utf-8"?>\\
<manifest xmlns:android="http://schemas.android\newline.com/apk/res/android" \\ package="com.example.myapp">\\
    <application>\\
    \textbf{<meta-data android:name="1232rt" android:value="100.50.43.22"/>}\\
        <activity ...>\\ 
        </activity>
    </application>
</manifest>}
    \caption{The IP address string 100.50.43.22 is transferred to the Manifest file, and is attributed a meta-data tag. The name of the meta-data tag replaces the string name in the former smali file (which needs a number of additional functions that are not mentioned here).}
    \label{fig:manifest pockets}
\end{figure} 
\end{enumerate}
\begin{algorithm}[!]
\caption{Smali based attack function}\label{alg:att_smali}
\begin{algorithmic}[1]
\Procedure{SB\_Attack}{$Manifest,Smali...,S$}
\For{\texttt{$s \in S$
      s.t. $W_s>0$}} \label{lst:line:loop_smal}
\State $s \gets find\_line(s,Smali)$\label{lst:line:smali_find}
    \State
    $Enc \gets Encoding(s)$\label{lst:line:encode}
    \If{\textit{s} is an API call}\label{lst:line:api}
    \State
    $Smali \gets Reflection(Enc,Smali)$\label{lst:line:reflect}
    \EndIf
    \State
    $Manifest \gets Manifest\_Pockets(Enc,Manifest)$\label{lst:line:poc}
\EndFor
\State \textbf{return} $Manifest,Smali...$\label{lst:line:end5}
\EndProcedure
\end{algorithmic}
\end{algorithm}
Algo.~\ref{alg:att_smali} describes the attack vector procedure, which implements the following steps:
\begin{enumerate}
    \item For each observation $s$ with $W_s$ > 0, run steps 3-8 (line~\ref{lst:line:loop_smal}).

        \item Find the observation in the smali code files (line~\ref{lst:line:smali_find}).
        \item Encode it with base64  encoding (line~\ref{lst:line:encode}).
        \item If it is an API call (line~\ref{lst:line:api}), use reflection (line~\ref{lst:line:reflect}).
        \item Use the Manifest\_Pockets method on the encoded string (line~\ref{lst:line:poc}).
        \item Return the manipulated files and all the other depackaged files as output (line~\ref{lst:line:end5}).
\end{enumerate}

\item \textbf{Combined attack (MA,1,1):} This attack vector combines the smali based attack and the Manifest based attack type 1. The attacker scans the report from Drebin (full details in Section~\ref{drebin}) along with to the Manifest file and the smali code files. For the observations from the Manifest file (permissions and components) it uses \textbf{MB1\_ATTACK} procedure and for the observations from the smali code it uses the \textbf{SB\_ATTACK} procedure.
Algo.~\ref{alg:att_combined} describes the attack vector procedure, which implements the following steps:
\begin{enumerate}
    \item Split the observations from Drebin into two sets - \textit{M\_observations} which is a set of observations of Manifest file and \textit{S\_observations} which are observations of Smali code files (line~\ref{lst:line:split_obs}).
        \item Run the \textbf{Smali\_attack} function on the observations and the smali codes. Update the file (line~\ref{lst:line:smal_obs}).
       \item Run the \textbf{MB1\_attack} function on the observations and the Manifest file. Update the files (line~\ref{lst:line:man_obs}).
        \item Return the manipulated files and all the other depackaged files as output (line~\ref{lst:line:end6}).
\end{enumerate}

\begin{algorithm}[!]
\caption{combined attack function}\label{alg:att_combined}
\begin{algorithmic}[1]
\Procedure{Combined\_Attack}{$Manifest,Smali...,S$}

\State $M\_observations,S\_observations \gets Split\_Observations(S)$ \label{lst:line:split_obs}
\State $Manifest,Smali... \gets SB\_Attack(Manifest,Smali...,S\_observations$)
\label{lst:line:smal_obs}

\State $Manifest,Smali... \gets MB1\_Attack(Manifest,Smali...,M\_observations$)
\label{lst:line:man_obs}

\State \textbf{return} $Manifest,Smali...$\label{lst:line:end6}
\EndProcedure
\end{algorithmic}
\end{algorithm}
\end{enumerate}
Table~\ref{attack_vector_summary} summarizes the attack vectors and their attack tuples.

\begin{table}[!h]
	\centering
	\caption{Attack vectors}
	\label{attack_vector_summary}
	\begin{tabular}{|c|c|c|c|}
		\hline
		Attack vector& Attack model &Smali access & Manifest access    \\ \hline
		MB1& MA&0& 1    \\	\hline
		MB2& MA&0& 1    \\	\hline
		MB3& DA&0& 1    \\	\hline
		MB4& ZK&0& 1    \\	\hline
		SB& MA&1& 0    \\	\hline
		Combined& MA&1& 1    \\	\hline
	\end{tabular}
\end{table}

\section{Android Malware Detection Systems}
In this section, we describe four Android malware detection systems: Drebin~\cite{arp2014drebin}, Kirin~\cite{enck2009lightweight}, Permission-Based Android Malware Detection (PB-AMD)~\cite{aung2013permission}, and FAMOUS~\cite{kumar2018famous}. These four detection systems represent four types of Android malware detection systems involving requested permissions. Drebin has several features one of which is requested permissions. Kirin, PB-AMD and FAMOUS use permission requests as their only feature. Kirin is a rule based detection system, whereas PB-AMD is a ML detection system. FAMOUS scores permission requests as a function of their frequency in benign and malicious apps, and calculates the maliciousness of an app based on these scores. Since Kirin, PB-AMD and FAMOUS only extract the permission requests as their feature set, their detection system fail when permission requests are concealed. Experimenting on these four different types of permission based detection systems led us to the conclusion that there is an ominous vulnerability in each one of them.
First, we describe the vulnerability. Then, we discuss each detection machine we used to assess our evasion attacks.
\subsection{Vulnerability inspection}
\subsubsection{Androguard exploit} Drebin, Kirin, PB-AMD and FAMOUS have two components in common. The first is their feature extraction tool in that they all use Androguard~\cite{desnos2011androguard} or similar approaches to extract their features. In addition, these systems consider permission requests from the user as a key feature. Some systems consider the permission requests as the only feature. Therefore, taping Androguard's feature extraction methods, we found that Androguard fetches permission requests by parsing the Manifest file. It searches for \textbf{uses-permission} tags and lists the permission constant strings they encloses. Any change to the structure of the uses-permission tag name or constant permission strings causes Androguard to skip the permission request. The permission request is concealed from the detection system. Therefore, we replaced the uses-permission name tag with the uses-permission-sdk-23 name tag. 

\subsubsection{Permission based detection exploit}
A malware detection system is based on comparing the vectors of values of benign and malicious apps in the train data. Therefore, a value that does not occur in any vector from the train data may result in wrong detection. In this study, we found that there are no use of permission group superior names in the train data. Therefore, we exploited this insight in our evasion attack and used superior permission group names instead of the regular constant names. For example, we replaced android.permission.READ\_SMS permission constant with its superior permission group constant android.permission.SMS. 

\subsubsection{Mixture of the two exploits}
A combination of this two phenomena produces the following vulnerability:
\begin{enumerate}
    \item Replace the uses-permission name tag with the uses-permission-sdk-23 name tag.
    \item Replace the permission constant with its superior group name.
\end{enumerate}

\subsection{Drebin}  
\label{drebin}
Drebin is a lightweight method for the detection of Android malware introduced by Daniel Arp in 2014. In this section, we only describe the full run of the Drebin classifier. The details on the features Drebin observes can be found in Section~\ref{drebin_reports}. 
Overall, the Drebin classification has the following form:
\begin{enumerate}
    \item Drebin saves the original labels of each app. A label of -1 for is used for benign apps, and 1 for malicious apps. It saves the original labels of each app.
    \item During the training process, Drebin inspects each training sample (APK file). It assigns each observation a corresponding weight $W_s$ and sums these weights.
    \item Using SVM, Drebin computes the maximum weight that corresponds to a benign app as $t$.
    \item In the test phase, if $\Sigma W_s > t$, Drebin predicts a label of 1 (malicious). Otherwise, it predicts the label of -1. It saves the weights and labels in \textit{Explanations.json} (see Fig. \ref{fig:rep} for an example).
    \item Drebin compares its predicted labels to the original labels of the apps, and produces an accuracy rate.
\end{enumerate}  
We used an implementation of Drebin that we found on Github~\cite{Drebin_implementation}. This version uses Androguard, as do all the other detection machines in this section.

\subsection{Kirin}
Kirin~\cite{enck2009lightweight} is a rule-based security system for Android introduced by William Enck in 2009. It does not use ML to detect Android malware. As this code integrates with the Andorid OS, changes to the operation system suggest that Kirin is not likely to work well with newer versions of Android. One of the original authors suggested using AXMLPrinter2 or Androguard to re-implement Kirin. Kirin inspects the use of Android permissions of an app against a list of rules at install time. The rules implemented in this system are:
 \begin{enumerate}
     
 \item An application must not have the SET\_DEBUG\_APP permission label.
\item An application must not have PHONE\_STATE, RECORD\_AUDIO, and INTERNET permission labels.
\item An application must not have PROCESS\_OUTGOING\_CALL, RECORD\newline\_AUDIO, and INTERNET permission labels.
\item An application must not have ACCESS\_FINE\_LOCATION, INTERNET, and RECEIVE\_BOOT\_COMPLETE permission labels.
\item An application must not have ACCESS\_COARSE\_LOCATION, INTERNET, and RECEIVE\_BOOT\_COMPLETE permission labels.
\item An application must not have RECEIVE\_SMS and WRITE\_SMS permission labels.
\item An application must not have SEND\_SMS and WRITE\_SMS permission labels.
\item An application must not have INSTALL\_SHORTCUT and UNINSTALL\_\\SHORTCUT permission labels.
\item An application must not have the SET\_PREFERRED\_APPLICATION permission label and receive Intents for the CALL action string
 
 \end{enumerate}
\subsection{Permission-Based Android Malware Detection} 
Permission-Based Android Malware Detection~\cite{aung2013permission} (which we termed PB-AMD) was introduced by Zarni Aung in 2013. This system solves an  unsupervised machine learning problem: can an unsupervised learning machine decide the correct label for an APK based solely on its requested permissions? The first step is to extract the permission requests.  As there is no available official implementation of this detection system, we implemented the feature extraction by Androguard.  Then, Information Gain~\cite{kent1983information} is used as their feature selection algorithm to reduce the feature set size. The authors suggested using K-means~\cite{wagstaff2001constrained,krishna1999genetic} to cluster the data into benign and malicious clusters based on the permission requests. Then, decision trees are used on each cluster. The decision tree classifiers the authors used were j48~\cite{bhargava2013decision,ruggieri2002efficient}, RF, CART~\cite{denison1998bayesian,steinberg2009cart}. The most promising algorithm in the original study was found to be RF, which was confirmed by our runs.
\subsection{FAMOUS}
FAMOUS (Forensic Analysis of Mobile devices Using Scoring
of application permissions)~\cite{kumar2018famous}
introduced APK permission scoring in 2018 and was developed by Dr. Ajit Kumar. The authors used the scoring of permission requests to detect malicious and benign apps. First, they counted occurrences of permission requests in benign apps (PuB) and in malicious apps (PuM). Then, they calculated
$MSP=\frac{PuM}{M}$ and
$BSP=\frac{PuB}{B}$ (where B is the number of benign apps, and M is the number of malicious apps).
The effective Maliciousness Score of Permission (EMSP) was calculated as the final score of the permissions: EMSP = MSP - BSP. The authors defined the Maliciousness Score of each permission $p_{i}$ by:
MS=$\Sigma_{i=1}^nEMSP(p_i)$. In total, the feature set the authors used included EMSPs and MS.
The authors used ML algorithms to predict malicious and benign labels for new apps. They implemented RF, SVM, KNN, NB, and CART, where the best performance was achieved by RF.

\section{Results}
\label{results}
We evaluated the effectiveness of our attack using the metrics from Section~\ref{eval}. We ran our assessment on an Intel(R) Core(TM) i7-4510U CPU with 2 GB RAM with GeForce 840M GPU. The Androguard version we used (version 3.3.5) is the latest commited version~\cite{desnos2011androguard}. We used $\sim$75K benign apps and $\sim$32K malicious apps (for more details on the source of our data, see Section~\ref{eval}). To generate a feasible ratio of benign to malicious files~\cite{pendlebury2018tesseract}, 
we used a ratio of 90/10. As we had more malicious apps than the one-tenth of the benign apps', we randomly split the malicious data into 5 parts and use the whole benign dataset with each part, separately. 

In addition, we used a ratio of 80/20 between the train and test data. We did not use the benign samples for the test data, since the goal was to assess our attack's evasion rate. 
Overall, we used $\sim 60k$ benign apps and $\sim 25k$ malicious apps as train data, and $\sim 6k$ malicious apps as test data. 
\subsection{VT verification}
\subsubsection{Benign apps}
We validated that our benign apps were identified as malicious by fewer than 3 VT scanners, as described in Section~\ref{eval}. We found that 92\% of the apps were not identified by any scanner, and 5\% were identified by 1 scanner. An additional 0.5\% of the applications were identified by 2 scanners. Hence, the benign dataset was verified as benign.
\subsubsection{Malicious apps}
We validated that our malicious apps were identified by at least 3 VT scanners, as described in Section~\ref{eval}. We found that only 0.03\% of the apps were not identified by any scanner, and 0.01\% were identified by 1 scanner. No apps were identified by 2 scanners. Hence, our malicious dataset was considered to be malicious since only 0.04\% of the apps were identified by the 3 scanners. 

\subsection{Evasion Robustness}
To get a clearer view of the evasion robustness for each evasion attack, we evaluated seven cases:
\begin{enumerate}
    \item Original malicious apps
    \item MB1 attack
    \item MB2 attack
    \item MB3 attack
    \item MB4 attack
    \item SB attack
    \item Combined attack
\end{enumerate}
PB-AMD and FAMOUS included multiple algorithms. Because we tested various numbers of features and several algorithms, we could not properly express all the options. Therefore, we picked the algorithm with the best performance we examined which was Random Forest in both detection machines (with 100 features in PB-AMD. FAMOUS used all the default Android permissions in its implementation). 

An app that triggered an error in the evasion attacks' construction was not included in the manipulated test data. Some of the errors we encountered were a result of corrupted APKs that we could not depackage or repackage. Other errors were a consequence of faults in the manipulation process. Overall, the error rate did not exceed 10\% of the test data.  

As can be seen in Table \ref{det_rate}, the Manifest Based evasion attacks decreased all the detection systems. Drebin sustained a 16\% evasion robustness despite our MB1 attack, and a fair evasion robustness of at least 70\%
with the other MB attacks. The other machines had an evasion robustness of 0\%. Since Drebin was the only machine that extracted features outside the Manifest file, its detection rate did not go to zero when confronting our manifest based attacks. Drebin's decrease in evasion robustness in the presence of MB1 manipulated apps may be due to the fact that in this specific manifest based attack, the features we conceal were components and requested permissions, unlike the other MB attacks which focus solely on permissions. As the SB attack did not change any permission requests, the evasion robustness rates of each machine remained similar to the original apps. The combined attack associated the manipulation of both the Manifest file and smali code files. Therefore, it affected the permission based detection machines in a similar way as the MB attack. Drebin, because it is based on both the Manifest and smali files, curtailed its greatest loss in all 6 attacks to 1\%.

\begin{table}[!h]
	\centering
	\caption{Evasion Robustness of Kirin, Drebin and PB-AMD for each case. }
	\label{det_rate}
	\begin{tabular}{|c|c|c|c|c|c|c|c|}
		\hline
		& Number of applications &Drebin (std) & Kirin  (std) &  PB-AMD-RF (std) & FAMOUS-RF (std)     \\ \hline
		Original&6327 & 0.964 (0.009)& 0.478 (0.004) & 	0.872 (0.009)&	0.856 (0.008)\\
		\hline
		
		MB1 attack&5808&  0.162 (0.015)&0 (0)  & 0 (0)& 0 (0) \\
		\hline
		MB2 attack&5864 & 0.71 (0.037)&0 (0)  & 0 (0)& 0 (0)\\
		\hline
		MB3 attack&5850& 0.784 (0.026) &0 (0)  &0 (0)  & 0 (0)\\
		\hline
		MB4 attack&5863&  0.798 (0.024)&0 (0)  & 0 (0)& 0 (0) \\
		\hline
		SB attack&5817 & 0.84 (0.012)& 0.478 (0.004) & 	0.872 (0.009)& 0.878 (0.01)\\
		\hline
		Combined attack& 5769 & 0.01 (0)&0 (0)  & 0 (0)& 0 (0) \\
		\hline
	\end{tabular}
\end{table}
\subsection{Functionally and maliciousness tests}
We then tested whether our evasion attacks would damages the functionality and maliciousness of the previous Android malware. An evasion attack is a risky and complex approach.  The resulting app may not harm the user, because it crashes in the initial steps of the app. We implemented the functionality test for our apps in an emulator using Pixel 2 XL image and SDK 27. We implemented a functionality test on each app before each evasion attack (see Section \ref{attacks}) and after it. For each app, our functionality test was implemented as follows:
\begin{enumerate}
    \item Cleaning of the emulator's log.
    \item Installation of the app.
    \item Running the app for 1 second.
    \item Termination of the app.
    \item Uninstalling of the app.
    \item Inspection of the emulator's log for crashes.
\end{enumerate}

Not all the apps in our dataset were error-free. Some of the apps that we used were designed for an older SDK version. Other apps from our dataset were missing some vital components and therefore were not installed properly on the emulator. Other changes to the smali codes files and manifest file that we implemented in our evasion attacks resulted in various runtime errors. As a result, some of the apps crashed. Although a nonfunctional app does not attack the user that runs it, we implemented the maliciousness test (see Section~\ref{malic_metric}) on the evasion attack apps. In our runs, some of the manipulated apps were successfully installed but their corresponding original apps were not. Given that the maximum difference between the original apps group and the manipulated apps groups did not exceed 3.8\%, we ignored the difference. 
 
 The results of both the functionality and maliciousness tests are depicted in Table~\ref{Func_Malic}. It shows the number of apps that did not have any errors in the manipulation process, the number of successfully installed apps and the results for both functionality and maliciousness tests. The SD of the maliciousness test is indicated since there were numerous options for number of scanners, whereas the functionality test is binary. As can be seen, the original malicious apps had a 90\% functionality rate. This was the baseline for our evasion attacks. The evasion attacks differed in functionality rate. The MB2, MB3 and MB4 evasion attacks had a high functionality rate. In other words, over 80\% of the apps stayed functional after the manipulation. SB and the combined attack had a low functional rate. The maliciousness rate appeared to go in the opposite direction. The SB and combined evasion attacks had a -9.5 and -10.79 maliciousness rate. The MB2, MB3 and MB4 attacks achieved a maliciousness rate between -7.7 to -6.05. The MB1 evasion attack appeared to be effective in terms of detection rate, and had a fair functionality and maliciousness rates.
 To get a clearer view of the maliciousness rates for each evasion attack, Table~\ref{Malic_stat} presents the statistics of the maliciousness test. We show the rates for three cases. The first case describes the situation where the number of scanners decreased after the evasion attack on the apps. The second case illustrates the occurrence of an equal number of scanners before and after our evasion attack on the apps. The third case depicts a higher number of scanners after the evasion attack on the apps. The results seems similar in each case.

In conclusion, whereas our evasion attacks' apps were proved to be malicious and decreased the accuracy rate of the detection systems, some of the manipulated apps suffered considerable losses in terms of level of functionality. At the same time, the maliciousness rate differentiated between some of the attacks. However, the maliciousness rate of all the evasion attacks increased. In other words, each of our attacks sustained malicious activity.

\begin{table}[!h]
	\centering
	\caption{Functionality and Maliciousness tests}
	\label{Func_Malic}
	\begin{tabular}{|c|c|c|c|c|}
		\hline
		& Number of apps& Successfully installed apps &Functionality \%& Maliciousness test (SD)  \\ \hline
		Original&6087 & 4331 & 0.9 & -  \\
		\hline
		MB1 attack&5803 & 2728 & 0.56 & -7.4/62.5 (2.95) \\
		\hline
		MB2 attack&5860 & 4501 & 0.82 & -7.7/63.04 (3.26)\\
		\hline
		MB3 attack&5865 & 4501 & 0.82 &-6.2/63.51 (2.76)  \\
		\hline
		MB4 attack&5878 & 4491 & 0.82  &-6.05/63.55 (2.88)  \\
		\hline
		SB attack&5813 & 4470 & 0.17 & -9.5/62.64 (3.48) \\
		\hline
		Combined attack&5751 & 2721 & 0.24 & -10.79/64.44 (3.6)  \\
		\hline

	\end{tabular}
\end{table}

\begin{table}[!h]
	\centering
	\caption{Maliciousness test in percent }
	\label{Malic_stat}
	\begin{tabular}{|c|c|c|c|}
		\hline
		& Lower than before& Equal number of scanners &Higher than before \\ \hline
		MB1 attack&0.994 & 0.002 & 0.004 \\
		\hline
		MB2 attack&0.984 & 0.008 & 0.007\\
		\hline
		MB3 attack&0.988 &0.005  &0 \\
		\hline
		MB4 attack& 0.977&0.01  &0 \\
		\hline
		SB attack&0.998 & 0.0007 & 0.001 \\
		\hline
		Combined attack&0.999 & 0.0003 & 0.0007 \\
		\hline

	\end{tabular}
\end{table}

\section{Permission inspection}
\subsection{Permission levels}
\label{perms_inspection_dis}
\begin{figure}[!]
\begin{minipage}{.47\linewidth}
  \includegraphics[width=\linewidth]{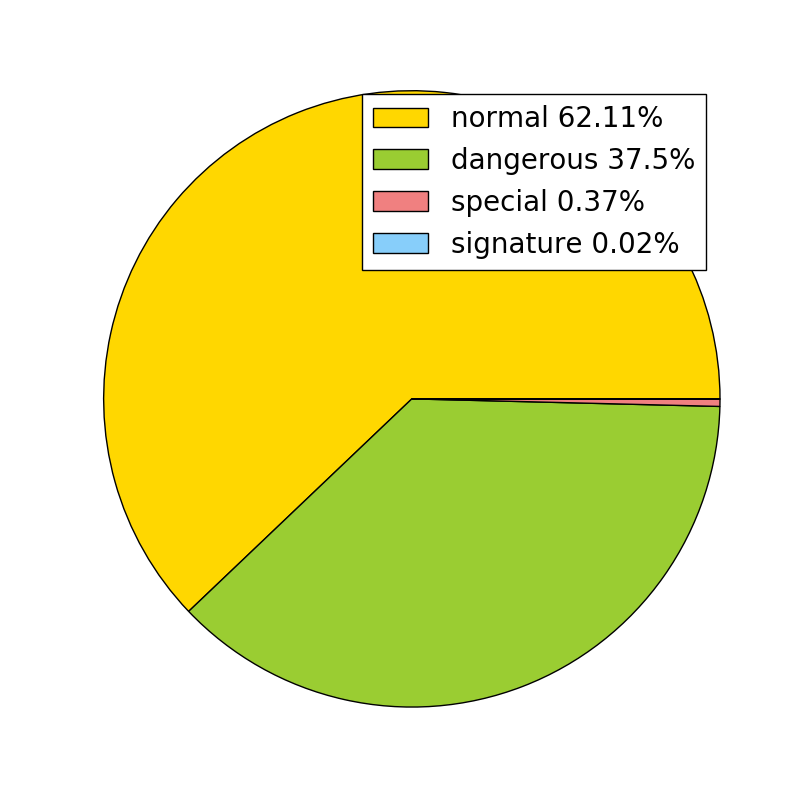}
  \caption{General protection level use, benign apps}
  \label{ben_sur}
\end{minipage}
\begin{minipage}{.47\linewidth}
  \includegraphics[width=\linewidth]{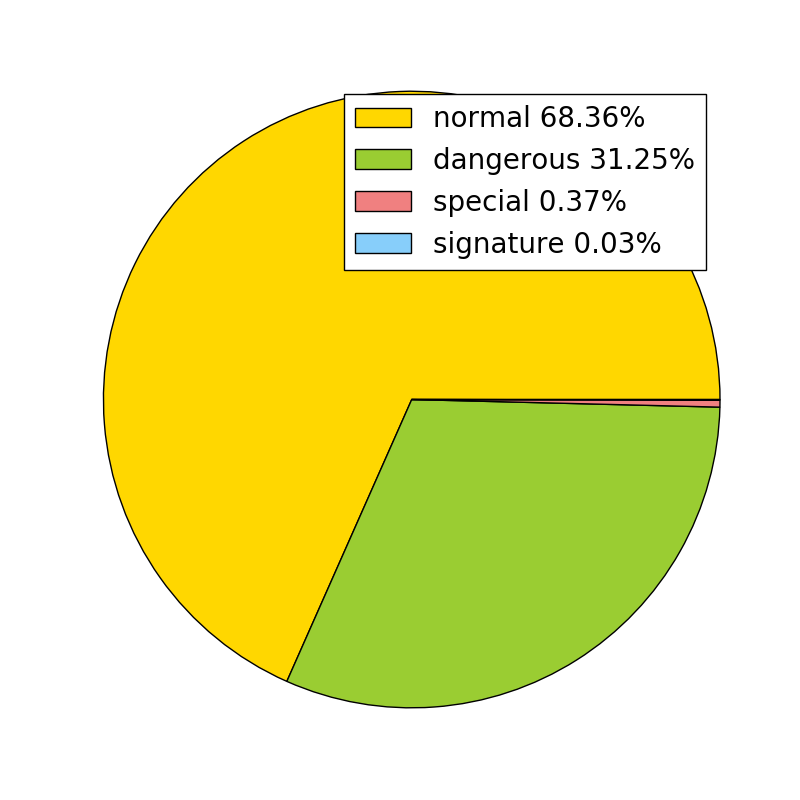}
  \caption{Dominant protection level, benign apps}
  \label{ben_dom}
\end{minipage}
\begin{minipage}{.47\linewidth}
  \includegraphics[width=\linewidth]{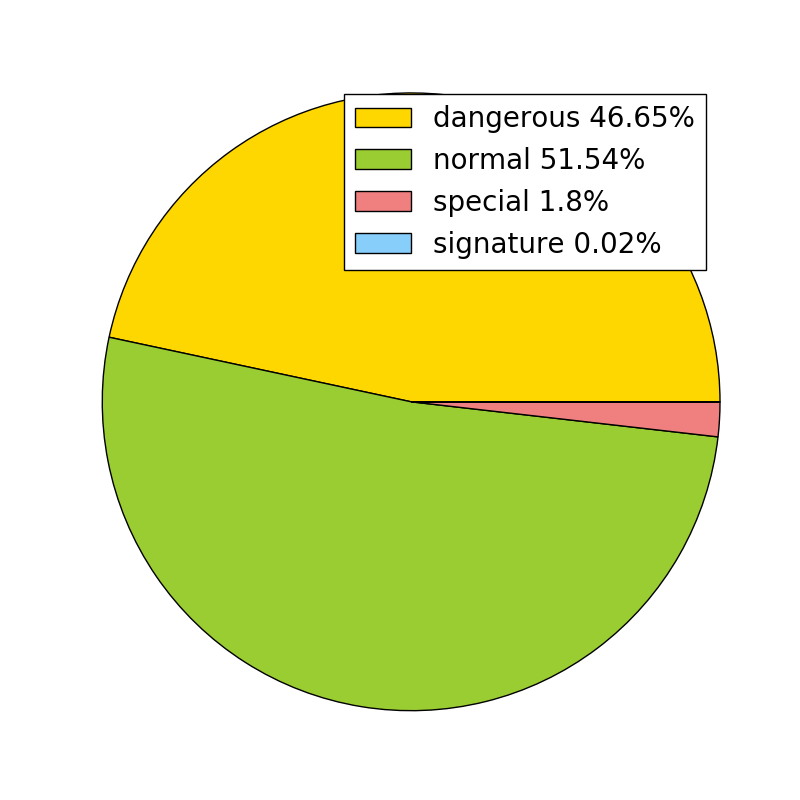}
  \caption{General protection level use, malicious apps}
  \label{mal_sur}
\end{minipage}
\begin{minipage}{.47\linewidth}
  \includegraphics[width=\linewidth]{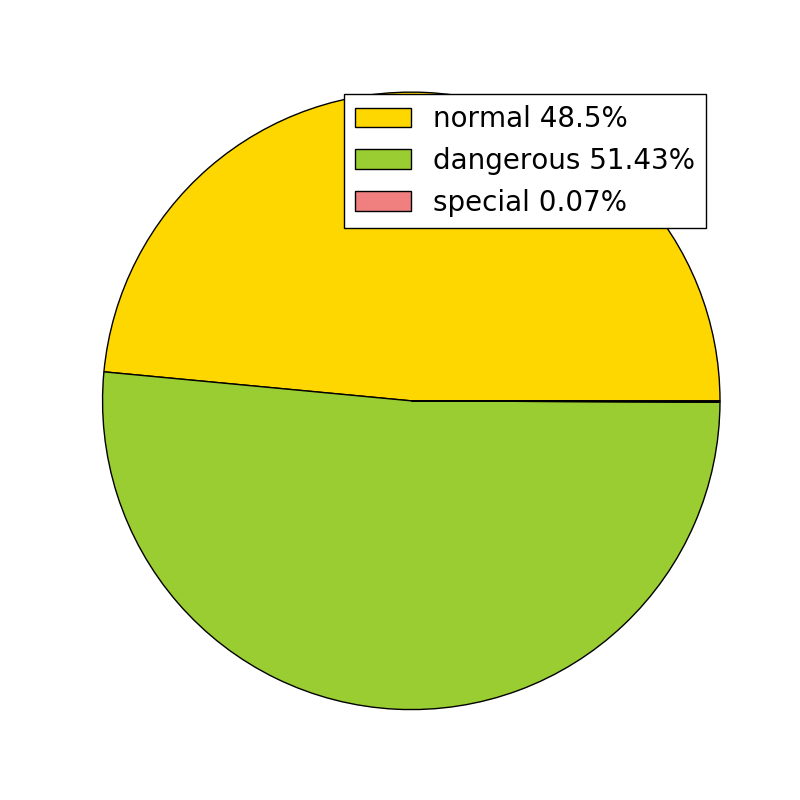}
  \caption{Dominant protection level, malicious apps}
  \label{mal_dom}
\end{minipage}
\caption{Pie diagrams of general/dominant protection levels.}
\label{pies}
\end{figure}

As mentioned in Section \ref{perms_ins}, we analyzed the behavior of the benign apps and malicious apps. We inspected the requested permissions of each group in general, and from the dominant permission point of view.
Fig. \ref{pies} depicts these behaviours. As shown in Fig. \ref{ben_sur}, more than 60\% of the benign apps fell into the normal protection level in the general protection level use. The dominant permissions (in Fig.~\ref{ben_dom}) had a stronger trend, $\sim70\%$ of the apps had a dominant permission from the normal protection level. The malicious apps group tended to have balanced behavior in both the general protection level use (Fig. \ref{mal_sur}) and the dominant protection level (Fig. \ref{mal_dom}). 

This observation suggests that a randomly typical benign app is likely to have more normal than dangerous permission requests in the manifest file. As a result, the dominant protection level in this kind of app is also normal. This insight can be exploited in the future in reverse: a malicious app's permission requests may not be dominated by a normal protection level.
\subsection{Permission Families}
As mentioned in Section~\ref{perms_ins}, we explored permission families. The permission family that only contains the android.permission.INTERNET permission was found in 78.51\% of the benign apps group and in 98.24\% of the malicious apps group. In other words, almost every malicious app used the internet permission, as did most of the benign apps. The explanation is straightforward since most of the apps in the world engage in network communication. Therefore, a malicious app that camouflage its permission requests such that only the Internet permission is detected by a permission based classifier is likely to be classified as a benign app. The first three families for both the benign and malicious apps are depicted in Tables~\ref{perm_fam_table_benign} and~\ref{perm_fam_table_malicious}. As can be seen, the top three permission families in the benign apps group were found to be INTERNET with 78.51\%, ACCESS\_NETWORK\_STATE with 58.6\% and INTERNET \& ACCESS\_NETWORK\_STATE with 58.26\%. The first three families in the malicious apps group were found to be INTERNET with 98.24\%, READ\_PHONE\_STATE with 95.52\% and INTERNET \& READ\_PHONE\_STATE with 95.4\% (the fourth and fifth families in this group are the second and third families from the benign apps group). It can be seen that the third family in each group is very close to the second family in terms of percentile, and is composed of the first and the second family. We concluded that in many benign apps, the Internet permission is connected to the network state permission. This is understandable since using internet is closely connected to the network state. In the malicious apps group the Internet permission appeared to be connected to the phone state. It is important for an attacker to get the phone number of the device, the status of any ongoing calls, etc. to evaluate the attack surface. 
\begin{table}[!b]
	\caption{Top permission families in the benign apps dataset.}
	\label{perm_fam_table_benign}
	\begin{tabular}{|c|c|}
		\hline
		"INTERNET"& 78.51\%\\
		\hline
    "ACCESS\_NETWORK\_STATE"& 58.6\%\\
    \hline
    "INTERNET"
    ,"ACCESS\_NETWORK\_STATE"& 58.26\%\\
    \hline

	\end{tabular}
\end{table}

\begin{table}[!]
	\caption{Top permission families in the malicious apps dataset}
	\label{perm_fam_table_malicious}
	
	\begin{tabular}{|c|c|}
		\hline
		"INTERNET"&98.24\%\\
			\hline
    "READ\_PHONE\_STATE"& 95.52\%\\
    	\hline
    "INTERNET"
    ,"READ\_PHONE\_STATE"&95.4\%\\
    \hline
	\end{tabular}
\end{table}

\section{Discussion and Conclusion}
Android malware detection is a popular research topic in security. Since the number of Android users is continuously increasing, the financial gain of an efficient and enhanced malware that evades detection has become highly attractive. Therefore, new malware is uploaded to various application markets every minute, taking advantage of unsuspecting users. As a countermeasure, a rising tide of detection and prevention systems are have been in both the academic and the industrial communities. 

In this study, we explored a new vulnerability in Androguard, a well known tool for the implementation of Android malware detection systems. This vulnerability can be exploited to decrease the detection rates of many detection systems that include requested permissions as a key feature. We showed that some detection systems drop to 0\% when an adversary exploits this vulnerability. Our future work will center on analyses of this type of detection systems.

In addition, we examined what defines a successful attack on an Android malware detection system. Is it merely the influence on the detection system's rate, or should we also consider other features such as the functionality and maliciousness of the manipulated app? Our conclusion is that functionality and maliciousness are extremely important when constructing evasion attacks. 
We suggest functionality and maliciousness tests to examine the functionality and malicious activity the app maintains after the manipulation.
\space To demonstrate, we proposed several novel evasion attacks that achieved different evasion rates. In addition, the maliciousness test showed that the manipulated apps maintained their high maliciousness value, since a greater number of VT scanners did not recognize the apps as malicious. However, the functionality test proved that the apps' functionality suffered a tremendous loss, from 90\% functional apps to 18\%-56\% functional apps in some cases. Other evasion attacks that we proposed maintained high functionality and maliciousness rates. Therefore, the impact of an evasion attack should be considered as a tradeoff between maliciousness and functionality rates. 
\section{Acknowledgements} 
We express our thanks to Dr. William Enck from the Department of Computer Science at NC State University for his insights on Kirin. We are grateful to Dr. Ajit Kumar from the Department of Computer Science at Sri Sri University for his FAMOUS implementation and his valuable assistance in running it. This work was supported by the Ariel Cyber Innovation Center in conjunction with the Israel National Cyber directorate of the Prime Minister's Office. 

%
%
%
\bibliographystyle{splncs04}
\bibliography{bibliography}

\end{document}